\documentclass[journal,dvipsnames,svgnames]{vgtc}                     % final (journal style)
\onlineid{0}

\vgtccategory{Research}

\vgtcpapertype{please specify}

\title{Vistrust: a Multidimensional Framework and \\ Empirical Study of Trust in Data Visualizations}

\author{%
  \authororcid{Hamza Elhamdadi}{0009-0006-8767-0681}, \authororcid{Adam Stefkovics}{0000-0003-4961-7792},
 \authororcid{Johanna Beyer}{0000-0002-3505-9171}, \authororcid{Eric Moerth}{0000-0003-1625-0146}, \authororcid{Hanspeter Pfister}{0000-0002-3620-2582}, \\\authororcid{Cindy Xiong Bearfield}{0000-0002-1451-4083}, \authororcid{Carolina Nobre}{0000-0002-2892-0509}
}

\authorfooter{
  \item
  	Hamza Elhamdadi and Cindy Xiong Bearfield are with UMass Amherst.
  	E-mails: helhamdadi@umass.edu, cindy.xiong@cs.umass.edu
  \item
  	Adam Stefkovics, Johanna Beyer and Hanspeter Pfister are with Harvard University
  	E-mails: adamstefkovics@fas.harvard.edu, jbeyer@g.harvard.edu, hpfister@g.harvard.edu 
    \item Eric Moerth is with Harvard Medical School
    E-mail: ericmoerth@g.harvard.edu
    \item Carolina Nobre is with the University of Toronto 
  	E-mail: cnobre@cs.toronto.edu
}

\abstract{%
 Trust is an essential aspect of data visualization, as it plays a crucial role in the interpretation and decision-making processes of users. While research in social sciences outlines the multi-dimensional factors that can play a role in trust formation, most data visualization trust researchers employ a single-item scale to measure trust. We address this gap by proposing a comprehensive, multidimensional conceptualization and operationalization of trust in visualization. We do this by applying general theories of trust from social sciences, as well as synthesizing and extending earlier work and factors identified by studies in the visualization field. We apply a two-dimensional approach to trust in visualization, to distinguish between cognitive and affective elements, as well as between visualization and data-specific trust antecedents. We use our framework to design and run a large crowd-sourced study to quantify the role of visual complexity in establishing trust in science visualizations. Our study provides empirical evidence for several aspects of our proposed theoretical framework, most notably the impact of cognition, affective responses, and individual differences when establishing trust in visualizations.   
}


\teaser{
  \centering
  \includegraphics[width=\linewidth]{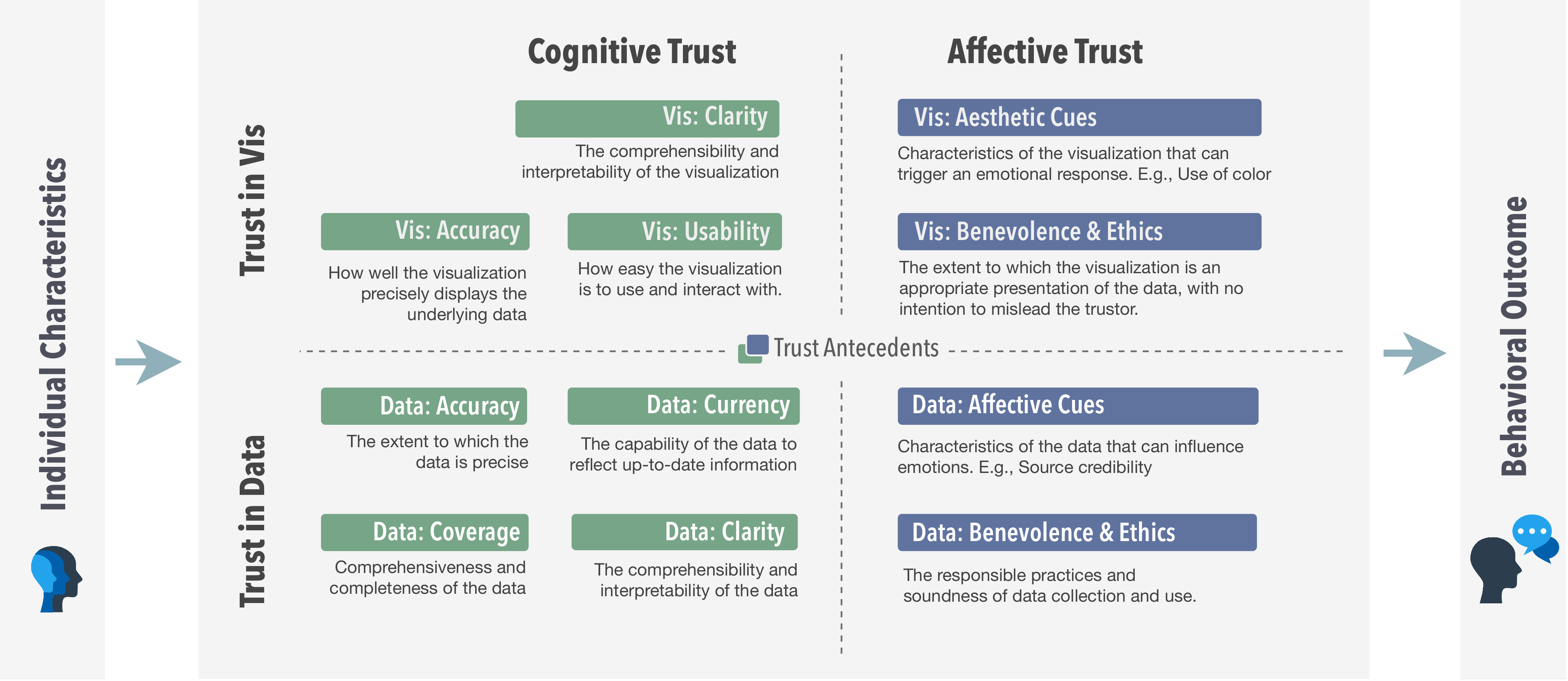}
  \caption{%
  	An integrated framework, which outlines the development of trust in visualizations. The framework defines the different \textit{trust antecedents} of the two basic components of trust (cognitive and affective trust). Both cognitive and affective trust can relate to the visualization and the underlying data. \textit{Individual characteristics} can play a role in shaping one's level of trust in visualizations, and \textit{behavioral outcomes} can emerge as a results of trust judgements.
  }
  \label{fig:teaser}
}

\graphicspath{{figs/}{figures/}{pictures/}{images/}{./}} % where to search for the images

\usepackage{tabu}                      % only used for the table example
\usepackage{booktabs}                  % only used for the table example
\usepackage{lipsum}                    % used to generate placeholder text
\usepackage{mwe}                       % used to generate placeholder figures

\usepackage{mathptmx}                  % use matching math font
\usepackage{colortbl}
\usepackage{enumitem}
\usepackage[dvipsnames,svgnames]{xcolor}
\definecolor{cognitive}{HTML}{D8EFE0}
\definecolor{affective}{HTML}{D6DFF2}
\definecolor{significant}{HTML}{D51B10}

\definecolor{significantEffect}{HTML}{D51B10}
\definecolor{trustColor}{HTML}{FFFFFF}
\definecolor{behavior}{HTML}{E0E0E0}

\definecolor{sequential1}{HTML}{08306B}
\definecolor{sequential2}{HTML}{08519C}
\definecolor{sequential3}{HTML}{2171B5}
\definecolor{sequential4}{HTML}{4292C6}
\definecolor{sequential5}{HTML}{6BAED6}
\definecolor{sequential6}{HTML}{9ECAE1}
\definecolor{sequential7}{HTML}{C6DBEF}
\definecolor{sequential8}{HTML}{DEEBF7}
\definecolor{sequential9}{HTML}{F7FBFF}

\newcommand{\pheading}[1]{\vspace{4px}\noindent\textbf{#1}}

\begin{document}

%%%%%%%%%%%%%%%%%%%%%%%%%%%%%%%%%%%%%%%%%%%%%%%%%%%%%%%%%%%%%%%%
%%%%%%%%%%%%%%%%%%%%%% START OF THE PAPER %%%%%%%%%%%%%%%%%%%%%%
%%%%%%%%%%%%%%%%%%%%%%%%%%%%%%%%%%%%%%%%%%%%%%%%%%%%%%%%%%%%%%%%

\firstsection{Introduction}
\maketitle

 As the field of data visualization matures and is more widely adopted in public settings, understanding the role of trust in visualizations becomes increasingly important, particularly when the data presented is urgent (e.g., climate change, Covid-19, etc.) For example, recent research exploring trust in Covid forecast visualizations showed how differences in visual encodings can significantly affect viewers' trust in the information and willingness to incorporate the information in their decision-making process \cite{padilla_multiple_2023}. 
 
 The concept of trust and the factors that can play a role in its formation has long been explored in the field of social sciences. From that research, two parallel paradigms of trust emerge: trust defined by Mayer et al. as ``the willingness of a party to be vulnerable to the actions of another party" ~\cite{mayer_integrative_1995} and trust defined by McAllister et al. as composed from cognition-based ``rational" and affect-based ``emotional" factors~\cite{mcallister_affect-_1995}. More specifically, \textit{cognitive trust} is defined as trust based on the knowledge and evidence of someone's ability and achievements, while \textit{affective trust} is defined as trust based on the emotional bond with someone \cite{mcallister_affect-_1995}. Research in this field also defines the factors that precede trust formation as \textit{antecedents} of trust. Examples of trust antecedents include benevolence and behavioral integrity \cite{mayer_integrative_1995}.  In this work, we build on the two-dimensional definition of trust as defined by McAllister and outline specific antecedents of both cognitive and affective trust in the context of visualizations (Figure \ref{fig:teaser}).
 
Researching trust in visualization has a fundamental difference from its application in the social sciences in that it does not concern itself with human-to-human relationships. Translating trust research from the social sciences to data visualizations can prove difficult because the trustee is a non-human subject \cite{mayr_trust_2019}. Nevertheless, several authors have argued that people may relate to non-human subjects socially ~\cite{kiesler_social_1997, nass_computers_1994} and are capable of trusting digital technology, information, or media ~\cite{kelton_trust_2007}. Furthermore, as shown by Kelton et al.~\cite{kelton_trust_2007}, trust in media can be interpreted on the same level as interpersonal trust (i.e., via a trustor-trustee relationship). 
 
 Recently, a handful of studies have attempted to measure trust in the context of data visualization ~\cite{kong_trust_2019, padilla_multiple_2023, kim_bayesian-assisted_2021, van_der_bles_communicating_2019, xiong_examining_2019}. Yet, these approaches typically focused on one or a few specific elements of trust, failing to capture the multi-faceted nature of trust in visualizations ~\cite{mayr_trust_2019}. Many of these approaches use a single-item scale that asks the user some variation of ``how much do you trust this visualization?" ~\cite{padilla_multiple_2023, kim_bayesian-assisted_2021, zhou_effects_2019, elhamdadi_using_2022}. This question is also asked without defining trust, instead requiring the participant to define trust for themselves before answering the question, which introduces error \cite{elhamdadi_how_2022}.  To improve how we measure and study trust in visualization, we propose a comprehensive, multidimensional conceptualization and operationalization of trust in visualization that builds on the trust literature and synthesizes existing work in the visualization field.

We operationalize this framework in a series of explicit measurements that cover cognitive and affective trust, as well as individual characteristics such as visual literacy, need for cognition, and demographic factors. We propose this framework to measure how a \textit{reader} forms trust in a visualization. This can be different from the trust formation of a \textit{creator}, who has the ability to manipulate visualizations and their underlying data. We run a large crowd-sourced study that serves as a case study for the framework and investigates the effect of visual complexity on establishing trust in science visualizations using the examples of Covid and crop disease visualizations. Our study provides empirical evidence for several aspects of our proposed trust framework. Among our most notable findings is the strong role that visualization topic has on trust mechanisms, and how complex visualizations can decrease trust and trigger more affective-based than cognitive-based trust judgments.

In this work, we make two main contributions: 
\begin{itemize}
    \item a multidimensional conceptualization
    and operationalization of trust in visualization. 
    \item an empirical study using this framework that investigates the role of visual complexity on cognitive and affective trust in science visualizations. 
    % \item study results that demonstrate how varying levels of complexity affect trust in science visualizations.
\end{itemize}

% There is currently a lack of consensus about the conceptualization of trust.
% Unidimensional vs. multidimensional
% Two paradigms:
% Trustworthiness (e.g. Mayer et al. 1995)
% Cognitive-affective trust (McAllister 1995)
% Attempts to synthesize (Tomlinson et al. 2020)
% Field of visualization – no clear definition of trust in visualization, earlier measurements focus only on partial element 

\vspace{-2mm}
\section{Related Work}

% summary statement for all approaches to trust measurement

Methods for measuring trust include uni-dimensional and multi-dimensional Likert scales, trust games, belief updating metrics, and processing fluency. Much research regarding these methods stems from the social sciences. Though, computer scientists and data visualization researchers have recently started implementing some of these methods to measure trust in our fields.

% high level summary statements (describe field/subfield) - what unifies, distinguishes subfields
% lots of different methods
% most are uni-dimensional likert scales, some belief updating methods

% the questions/items in each scale

% end each paragraph with a sentence on what they did vs what we do (and why)

% Hamza
%\subsection{Trust Measurement in Social Sciences}

%Social science trust research has formulated two primary theories of trust in parallel: trust as independent of trustworthiness~\cite{mayer1995integrative}, and trust as composed of cognition and affect~\cite{mcallister1995affect}.
\vspace{-1mm}
\subsection {Trust Measurement in Computer Science}
% Hamza
% How trust is measured across disciplines (scales,etc.)

Prior approaches to measuring trust in computer science research have used several different Likert scales that vary widely on the number and content of items included~\cite{kim_data_2018, kim_bayesian-assisted_2021, padilla_multiple_2023, jian_foundations_2000, liao_user_2022, xiong_examining_2019, artz_survey_2007} as well as the definition of trust they use \cite{artz_survey_2007}.
Some approaches simply use uni-dimensional scales that vary on the number of discrete values the user can choose (e.g., 0 to 100 \cite{kim_data_2018,padilla_multiple_2023}, 1 to 5 \cite{dasgupta_familiarity_2017,kim_bayesian-assisted_2021}). Some approaches~\cite{kim_data_2018, dasgupta_familiarity_2017} additionally ask participants to explain their reasoning via text response.
Other computer science researchers have used multi-dimensional Likert scales that vary along the number of dimensions and the number of discrete values allowed for the response \cite{gutzwiller_positive_2019, jian_foundations_2000, liao_user_2022, zhang_shifting_2022}.
A widely-used 12-item scale, developed by Jian et al.~\cite{jian_foundations_2000}, captures user trust in specific automated systems, as opposed to previous scales from social sciences that aimed to measure general trust in automated systems. 
The user is asked to rate each of the items, which are ordered from negative to positive valence (e.g., The system is deceptive, ..., The system is dependable), from 1 to 7. 
However, the ordering of the items in this scale was shown to significantly bias users towards positive ratings when compared to both a randomized sequence of the twelve items and a flipped version where the items were ranked from positive to negative valence~\cite{gutzwiller_positive_2019}.

A more complex method for measuring trust seen in computer science research involves measuring the likelihood and frequency of a user to update their initial beliefs based on the recommendation and accuracy of a machine learning model \cite{yin_understanding_2019, yu_trust_2016, yu_user_2017}. The use of this approach to measure trust is validated by findings including users of machine learning systems can determine the accuracy over successive uses of a system \cite{yu_trust_2016}, users' trust in a system is more heavily impacted by its failures than successes \cite{yu_user_2017}, and users are impacted by a system's response as well as its stated accuracy \cite{yin_understanding_2019}.

Few computer science trust researchers have employed the use of trust games. As far as we are aware, from the non-visualization computer science trust research corpus, only Zheng et al.~\cite{zheng_trust_2002} have measured the impact of technology-mediated communication on trust using an investment game. The goal of this research was to determine how meeting a person via an online social chat room affected trust when compared with meeting face-to-face, or not meeting at all. This study, therefore, shares more in common with existing trust game research in social science than with the hypothetical trust game approach proposed by Elhamdadi et al. ~\cite{elhamdadi_how_2022} in which one of the participants would be a visualization or an automated system.

\vspace{-1mm}
\subsection {Trust Measurement in Data Visualization}

Measuring trust in data visualization research has been historically inconsistent and highly variable \cite{kim_data_2018, padilla_multiple_2023, zehrung_vis_2021, kong_trust_2019, xiong_examining_2019, zhou_effects_2019, kim_designing_2020}. Many data visualization trust researchers use a single-item scale to measure trust \cite{padilla_multiple_2023, kim_data_2018, dasgupta_familiarity_2017, kim_bayesian-assisted_2021, zhou_visual_1998}. These scales ask users to rate their trust (or agreement with statements that measure trust) on a scale with varying discrete value ranges (e.g. ``how trustworthy you think the graph is as a whole" on a scale from 0 to 100 \cite{padilla_multiple_2023}, ``how likely [you] thought it was that the data was manipulated" on a scale from 1 to 5 ~\cite{kim_bayesian-assisted_2021}, ``rate the trust level of predictions on which decisions were made" on a scale from 1 to 9 ~\cite{zhou_effects_2019}, etc.), However, single-item scales are insufficient to capture trust ~\cite{evans_survey_2008} due to the multi-faceted nature of trust ~\cite{kelton_trust_2007}. Furthermore, these uni-dimensional scales do not use the same number of discrete values, nor do they provide an explicit definition of trust, which can cast doubt on the validity of these approaches. Conversely, our framework measures trust with multi-item scales that capture trust and its antecedents, as described in established social science research ~\cite{mcallister_affect-_1995}. We also consider an existing multi-item trust measurement scale ~\cite{xiong_examining_2019} from data visualization research that uses three components of transparency (i.e., accuracy, clarity, and disclosure ~\cite{schnackenberg_organizational_2016}) as a proxy to measure trust. Participants were asked to rate each of the components of trust on a scale from 1 to 5; the results of this work validated the known positive correlation between transparency and interpersonal trust \cite{jahansoozi_organization-stakeholder_2006} from existing social sciences research in the context of data visualizations ~\cite{xiong_examining_2019}. 
% Therefore, our multi-item trust scale also includes items regarding accuracy, clarity, and disclosure. 

Recent trust research in visualization also indicates that processing fluency, the ease with which one interprets a stimulus, may function as a surrogate measure of trust in data visualizations \cite{elhamdadi_how_2022, elhamdadi_using_2022}, which follows from prior social and computer science research that demonstrates a link between fluency and trust \cite{thorndike_constant_1920, wetzel_halo_1981, sohn_consumer_2017}. High-intensity affective responses have also been linked to positive experiences of ease viewing a stimulus (fluency) \cite{landwehr_nature_2020}. Hence, the affective trust items from our framework indirectly capture participants' experience of fluency and, by proxy, trust.

% The recent finding that a single visualization is enough for some people to update their prior beliefs \cite{xiong2022seeing} indicated that measuring the likelihood of participants to update their prior beliefs (called belief updating) may be a method of measuring trust. This hypothesis was tested by Kim et al.~\cite{kim_bayesian-assisted_2021}, where they modeled the expected change in initial beliefs via a Bayesian network based on the intensity of the claims made by the visualization and compared with the post-study beliefs. 
% Similar to Yin et al.~\cite{yin2019understanding}, Kim et al.~\cite{kim_bayesian-assisted_2021} 
% %As in \cite{yin2019understanding}, \cite{kim_bayesian-assisted_2021} 
% found that a change in the initial beliefs to agree with the data shown in a visualization is correlated with higher trust. Our method for measuring trust captures participants' affective trust at the beginning of our study and compares this with their cognitive trust after interacting with the visualization.

While Elhamdadi et al.~\cite{elhamdadi_how_2022} argue that trust games, particularly investment games, can be used to measure trust in visualizations, visualizations present a difficult confound to overcome when using investment games. In a standard investment game~\cite{berg_trust_1995}, the reasoning for one participant's decision to invest a certain amount is not known to the other participant. However, when using a visualization, the underlying data becomes a confound. The visualization investment game, as discussed by Elhamdadi et al.~\cite{elhamdadi_using_2022}, does not account for the assessment of the underlying data in their decision to invest, thus confusing the correlation between investment and trust in the visualization. Future research into trust games in visualizations may ameliorate this issue.

\section {An Integrated Framework for Trust in Visualization}
\label{sec:trust_framework}
We propose a framework to comprehensively interpret and measure trust in visualizations by applying general theories of trust \cite{mayer_integrative_1995, mcallister_affect-_1995}, and by synthesizing and extending earlier work and factors identified by previous studies in the visualization field \cite{kelton_trust_2007, kong_trust_2019, mayr_trust_2019, van_der_bles_communicating_2019, van_der_bles_effects_2020, xiong_examining_2019}.

\noindent
\textbf{Uni- and Two-dimensional Theories of Trust}
Some approaches to trust measurement suggest that trust is uni-dimensional, many scholars argue that trust is multi-dimensional. The most prominent uni-dimensional understanding of trust was proposed by Mayer et al.~\cite{mayer_integrative_1995}. According to this approach, (1) trust is uni-dimensional around the key concept of “being vulnerable” and (2) trust should be distinguished from its antecedents. Mayer et al.~\cite{mayer_integrative_1995} identified ability, behavioral integrity, benevolence, and values congruence as the four antecedents of trust.

McAllister's two-dimensional conceptualization ~\cite{mcallister_affect-_1995} has been by far the most widespread theoretical basis for understanding trust. He defined interpersonal trust as “the extent to which a person is confident in and willing to act on the basis of, the words, actions, and decisions of another” \cite[pg. 25]{mcallister_affect-_1995}. Cognitive trust is grounded in judgments about the trustee’s perceived reliability and dependability. These quality judgments may serve as a rational basis for trust. Affect-based trust, on the other hand, consists of emotional bonds between individuals based on expectations of interpersonal care and concern. As Chua et al.~\cite{chua_head_2008} put it, cognitive and affective trust are “trust from the head” and “trust from the heart” respectively.
%~\cite{tomlinson_revisiting_2020}. % (cited in \cite{tomlinson_revisiting_2020}).

\noindent
\textbf{The Overlap in Cognition and Affect}
It is worth noting that the distinction between cognitive and affective trust antecedents is made at the experiential level \cite{duncan_affect_2007}. That is, a person assesses trust in a stimulus differently depending on whether they \textit{react} affectively or cognitively to aspects of the stimulus. Hence, in the context of visualizations, a visual cue is not objectively cognitive or affective but can be experienced by viewers in different ways. For example, someone can experience the clarity of a visualization as an aesthetic quality, leading to affective trust. On the other hand, they may experience clarity as a facilitator for interpreting the visualization, leading to cognitive trust. In this paper, we consider cognitive and affective trust antecedents as those that capture the viewer's cognitive and affective experiences with the visualization respectively.

\vspace{-1mm}
\subsection{Our Trust Framework}

Our framework relies on theoretical work by Tomlinson et al.~\cite{tomlinson_revisiting_2020}, who argues that the two paradigms (McAllister and Mayer) are not that distinct and aims to build bridges between them. First, we can consider the similarity of McAllister's assessment that trust “enables people to take risks” and the vulnerability concept from Mayer et al.~\cite{mayer_integrative_1995}. Second, like McAllister et al., Mayer et al.~\cite{mayer_integrative_1995} admit that trust is multifaceted and its antecedents are multi-dimensional. Tomlinson et al.~\cite{tomlinson_revisiting_2020} theorize that Mayer's antecedents can be related to McAllister's cognitive-affective framework; for instance, ability and behavioral integrity predict cognition-based trust, and benevolence and values congruence predict affect-based trust. Their findings indicate that cognitive and affective trust may have different  antecedents.

\pheading{Cognitive vs Affective} As depicted in Figure \ref{fig:teaser}, we apply the two-dimensional approach to trust in visualizations and characterize antecedents as \textit{cognitive} or \textit{affective}. Trust in visualization may be strongly determined by cognitive factors such as cognitive understanding; nevertheless, when cognitive trust is high, individuals may still distrust a visualization because of their negative emotions toward the visualization or its topic (or vice versa). This phenomenon can be attributed to the fact that feelings often arise with little or no cognition \cite{hoch_time-inconsistent_1991}.  For example, people may have negative feelings towards the data that is being visualized because it covers emotionally-negative material (e.g., Covid-19 deaths) or towards the visualization itself because some elements of the visualization (e.g., aesthetic elements) may induce negative feelings. Hence, these negative emotions may influence trust and decision-making regardless of cognition-based assessments.

\pheading{Data vs Visualization} We identified trust antecedents separately for the visualization and the underlying data as suggested by Mayr et al.~\cite{mayr_trust_2019}. Our framework includes 7 cognition-based and 4 affect-based antecedents of trust (see Fig \ref{fig:teaser})

\pheading{Individual Characteristics} Additionally, we argue that trust in visualizations may be subject to certain individual characteristics of the trustor. On the one hand, individual characteristics can alter the paths of trust development. For example, individuals with a higher need for cognition may weigh the cognitive aspects of a visualization higher, whereas others may be more likely to rely on affective responses to the visualization \cite{cacioppo_need_1982}. On the other hand, the likelihood to trust may differ between individuals regardless of the visualization’s characteristics. For instance, the Big Five personality traits have been linked to one's general propensity to trust \cite{evans_survey_2008}. Trust is positively associated with agreeableness, extraversion, and negative neuroticism, although these relationships are not deterministic and the relevance of personality traits in determining trust in visualizations remains unclear. Other known associations between human-to-human trust, cognitive skills \cite{sturgis_does_2010, hooghe_cognitive_2012}, the level of education \cite{borgonovi_relationship_2012, hooghe_cognitive_2012}, and political attitudes \cite{carlin_political_2018} have remained largely unexplored in data visualization research.

\pheading{Behavior} Lastly, our model addresses the consequences of trust in visualizations. Trust is considered a key criterion for engagement in information-related behavior \cite{kelton_trust_2007}. When trust is established, the trustor is willing to "take risks" \cite{mayer_integrative_1995}, i.e., integrate the information provided by the visualization into their actions. In turn, low levels of trust, or \textit{mistrust}, likely result in ignorance.

% \begin{figure}[h]
% \includegraphics[width=8cm]{figs/framework.png}
% \end{figure}

% Fig. x. Integrated framework for trust in visualizations
\vspace{-2mm}

\subsection {Antecedents of Cognitive Trust}
Cognition-based trust may originate from different rational evaluations. These assessments explain perception about the ability of the data or visualization to successfully describe their subject matter. We collected cognition-based antecedents by synthesizing previous studies in the visualization field \cite{kelton_trust_2007, kong_trust_2019, mayr_trust_2019, van_der_bles_communicating_2019, van_der_bles_effects_2020, xiong_examining_2019, zhang_affective_2013}. For each category, we provide data-driven examples from our survey that asked respondents an open question about why they did or did not trust a specific visualization.

\vspace{-2mm}

    \subsubsection*{Data Antecedents}
\paragraph*{Accuracy}
Accuracy refers to the extent to which the data is precise or free from error \cite{kelton_trust_2007, rieh_understanding_1998}. Such errors may originate, for example, from measurement error \textit{``The data is not entirely accurate, there were people who did not report having Covid, which made it difficult to keep an accurate count."} or lack thereof \textit{"The data falls right where I expect it to be."} Nevertheless, perceptions about the accuracy of the data can be a function of affective responses to the data source. Thus, accuracy may have both cognitive and affective antecedents. \textit{"One of the frames stated that the data came from the CDC. They are probably as close to accurate that we are going to get."} 

\vspace{-2mm}

\paragraph*{Currency}
Currency refers to the capability of the data to reflect up-to-date information. In some cases (e.g., rapidly-changing subjects or topics), this aspect can be crucial in determining the personal relevance of the information. \textit{``It is relevant, because it provides up-to-date Covid-data."}  
\vspace{-2mm}

\paragraph*{Coverage}
Coverage refers to the comprehensiveness and completeness of the data. Individuals may trust or distrust the visualization to the extent that it captures important aspects of the subject: \textit{``It has numbers that span over a period of time which shows a pretty big picture."}
\paragraph*{Clarity}
\vspace{-2mm}

Clarity refers to the comprehensibility and interpretability of the data \cite{van_der_bles_communicating_2019, van_der_bles_effects_2020}. The extent to which individuals understand the meaning of the data can influence trust regardless of how the data is actually presented. \textit{``The data makes sense to me."}
    \vspace{-2mm}

\subsubsection*{Visualization Antecedents}
\paragraph*{Accuracy}
The accuracy of the visualization refers to how well the visualization transparently and precisely displays all relevant elements of the underlying data. \textit{``The data seems to be presented in a complete manner, not leaving out any facts that could obscure the point presented."} Accuracy is also related to the extent to which the visualization captures all possible states and alternatives considering the underlying data \cite{xiong_examining_2019}. The accuracy dimension is particularly relevant in relation to uncertainty communication \cite{sacha_role_2016}. 
\paragraph*{Clarity}
Clarity of the visualization refers to the comprehensibility and interpretability of the visualization \cite{xiong_examining_2019}. Clear visualizations are easy to understand and do not contain obscure elements. \textit{``Everything is laid out in a way that is easy to see and understand."}
\paragraph*{Usability}
Usability refers to how easy the visualization is to use and interact with \cite{atoyan_trust_2006, costante_-line_2011, mayr_trust_2019}. For instance, Beauxis-Aussalet et al.~\cite{beauxis-aussalet_role_2021} argued that interactive visualizations are an “utmost important role in fostering trust in AI”. \textit{"Considering the actual data shows up over the mouse over, it provides more clarity".}

\vspace{-1mm}
\subsection {Antecedents of Affective Trust}
Previous research does not provide clear insights into the development of affective trust in human-non-human relationships. Affective trust is constituted of general emotional attachments between the two parties and the extent to which a trustee is believed to be good (and not cause harm) to the trustor \cite{mcallister_affect-_1995}. To apply the first part of the definition to visualizations, one can interpret affective trust as the immediate emotional responses to the data or visualization, as well as the affective perceptions of the data or visualization \cite{zhang_affective_2013}. Affective cues may include the objectivity or credibility of the data and/or the aesthetics of the visualization. Some research suggests that aesthetic judgments are driven mainly by affective responses \cite{bhandari_understanding_2019}. Notably, emotions toward the data may also originate from the emotions toward the source of the data \cite{kelton_trust_2007, mayr_trust_2019}. The second part of the definition resonates well with Mayer et al.'s notion of \textit{benevolence}\cite{mayer_integrative_1995}. In the case of visualizations, benevolence can be understood as the ethical use of the data and perceptions about the benevolent handling of the data and its visualization as free of any intention to cause harm to the trustor (e.g. by being intentionally misleading). Accordingly, we define affective cues and benevolence as separate antecedents of affective trust.
 
 \vspace{-2mm}

 \subsubsection*{Data Antecedents}
 \vspace{-1mm}
\paragraph*{Affective Cues of the Data} 

Affective cues are specific characteristics of the data that can influence emotions \cite{zhang_affective_2013}. For instance, these may include the perceived objectivity or believability of the data or the extent to which the data seems unbiased and credible. Negative affective trust in the data may also originate from a negative characteristic of the data source. \textit{``The visualization was presented with information from the CDC. I trust the CDC." "I just do not believe vaccinated vs. unvaccinated has any difference in cases."} 
\paragraph*{Benevolence and Ethics regarding the Data}
Benevolence and ethics regarding the data refer to the responsible practices and soundness of the data collection, the responsible use of data, and the appropriate citation of the data sources. The lack of information regarding these details may evoke a negative affect toward the data. \textit{``I have no idea how the data was gathered, if this is a 'dishonest' research study, etc. Where did the visualization come from? Who provided the data? How was it collected?"}

\vspace{-2mm}
 \subsubsection*{Visualization Antecedents}
\paragraph*{Aesthetic Cues of the Visualization}
Aesthetic cues of the visualization are specific characteristics of the visualization that can trigger an emotional response\cite{zhang_affective_2013}. These may involve specific use of colors, sizes, images, etc. \textit{"It looks scientific and seems professional." `` I don't really believe it, but I like that it's colorful."}
\paragraph*{Benevolence and Ethics regarding the Visualization}
Benevolence and ethics regarding the visualization refer to the responsible practices and accuracy of the visualization, the extent to which the visualization provides an appropriate presentation of the data, free of any intention to mislead the trustor. \textit{``I believe that charts like this misrepresent the truth and present things in an oversimplified way in order to mislead the viewer."}

\vspace{-2mm}

\section {Empirical Case Study}

We apply the trust framework in a case study investigating the factors that play a role in establishing trust in science-based data visualizations. More specifically, we are interested in determining the role of visual complexity in promoting trust. To capture the multi-dimensional aspects of trust in the study, we follow the two dichotomies described in our framework: the first between cognitive and affective trust elements and the second between trust in data and trust in visualization.  A detailed discussion on how we measure the different components of trust in the study can be found in Sections \ref{sec:affective_trust}, \ref{sec:trust_in_vis} and \ref{sec:trust_in_data}.

\vspace{-1mm}
\subsection {Study Design}
\label{sec:study_design}
We designed the study with three independent variables: visual complexity (simple, moderate, and complex), chart type (bar chart and line chart), and data topic (Covid Vaccines and Crop Diseases in Croatia). 
\begin{figure*}[!h]
     \centering
     \begin{subfigure}[b]{0.3\textwidth}
         \centering
         \includegraphics[height=4.5cm]{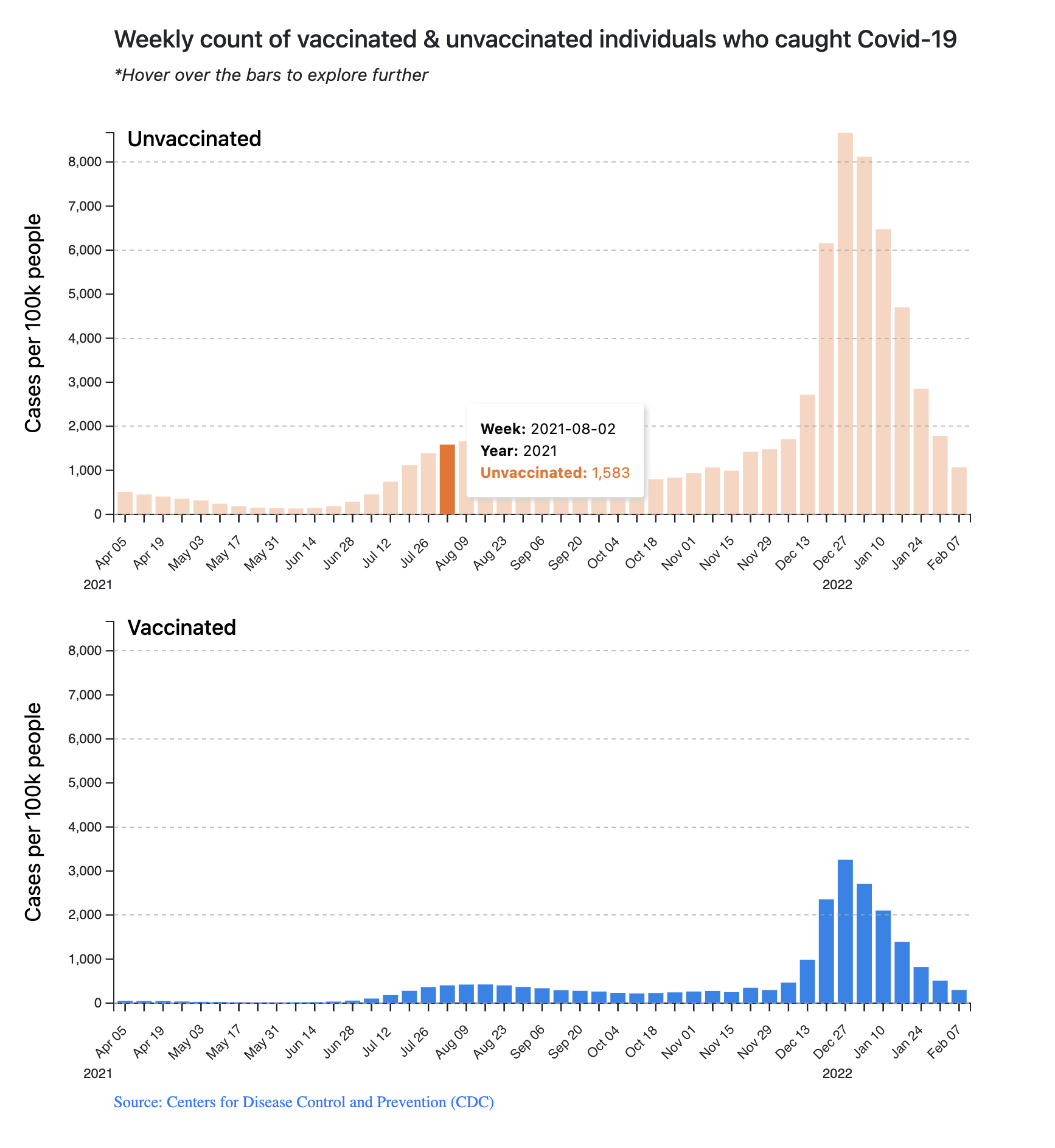}
         \caption{Bar chart of simple complexity.}
         \label{fig:simple_bar}
     \end{subfigure}
     % \hfill
     \begin{subfigure}[b]{0.3\textwidth}
         \centering
         \includegraphics[height=4.5cm]{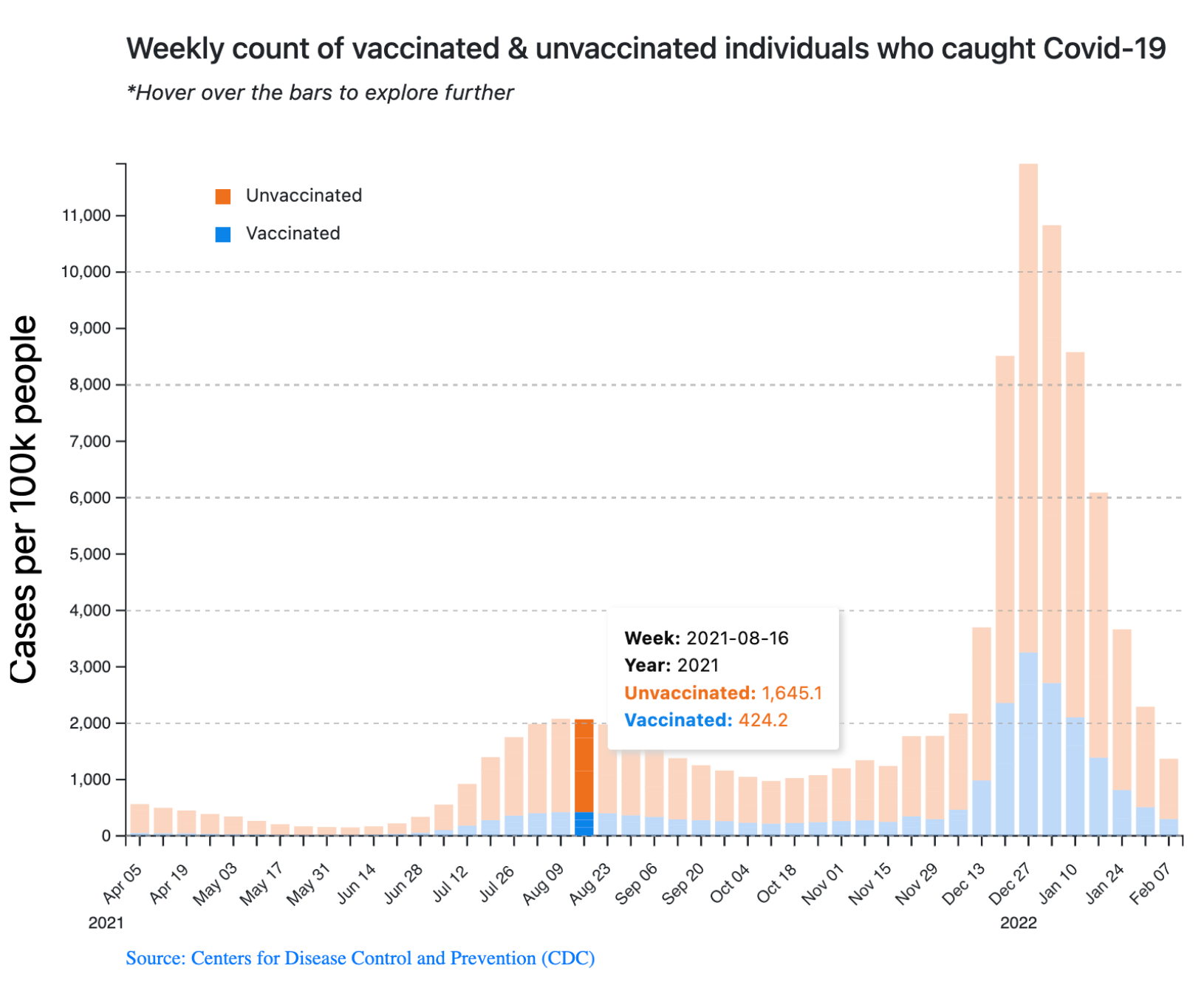}
         \caption{Bar chart of moderate complexity.}
         \label{fig:moderate_bar}
     \end{subfigure}
     % \hfill
     \begin{subfigure}[b]{0.3\textwidth}
         \centering
         \includegraphics[height=4.5cm]{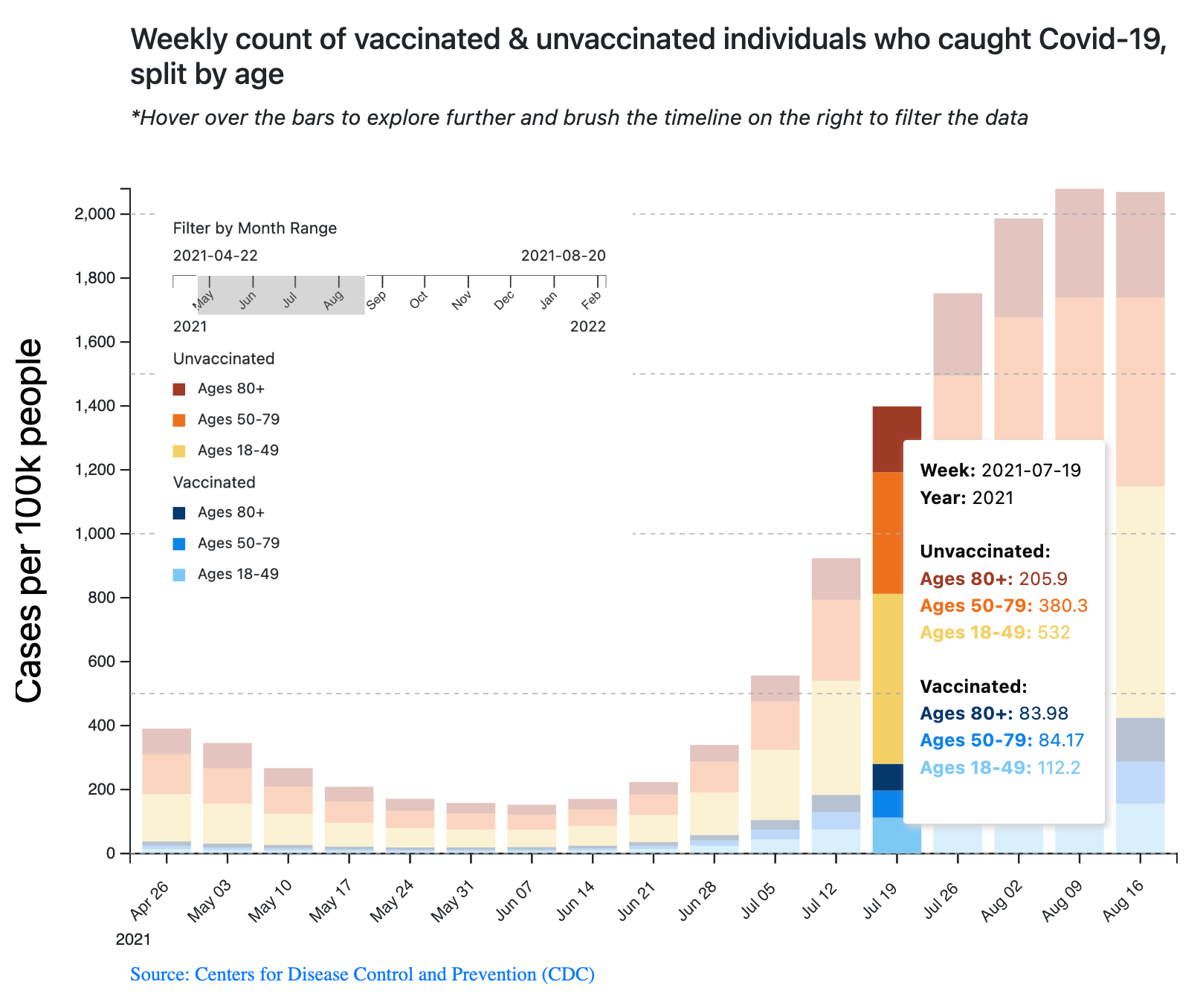}
         \caption{Bar chart of high complexity.}
         \label{fig:complex_bar}
     \end{subfigure}
\vspace{2mm}
          \begin{subfigure}[b]{0.3\textwidth}
         % \centering
         
         \includegraphics[height=4cm]{../figs/simpleLine.png}
         \caption{Line chart of simple complexity.}
         \label{fig:simple_line}
     \end{subfigure}
     % \hfill
     \begin{subfigure}[b]{0.3\textwidth}
         \centering
         \includegraphics[height=4cm]{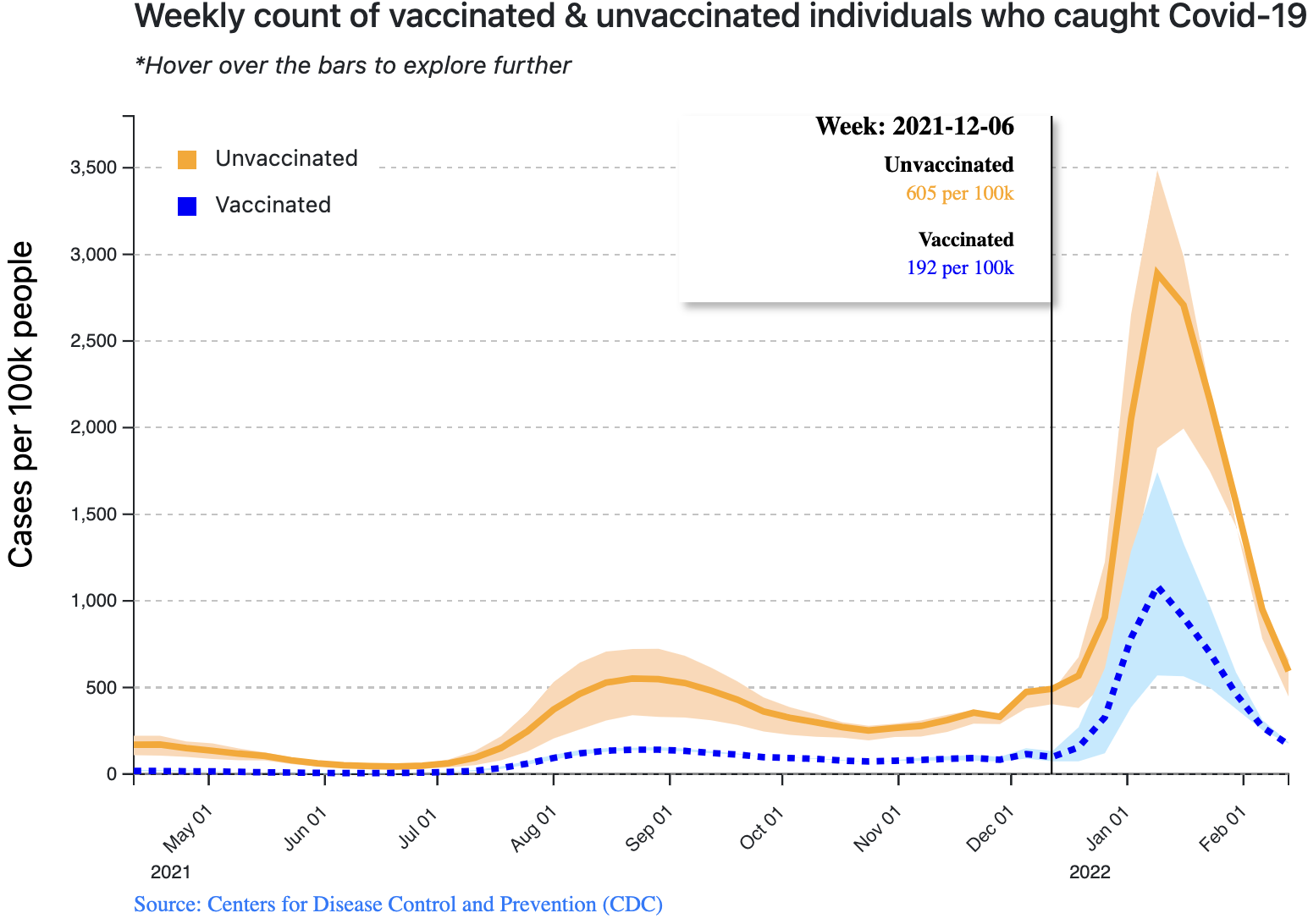}
         \caption{Line chart of moderate complexity.}
         \label{fig:moderate_line}
     \end{subfigure}
     % \hfill
     \begin{subfigure}[b]{0.3\textwidth}
         \centering
         \includegraphics[height=4cm]{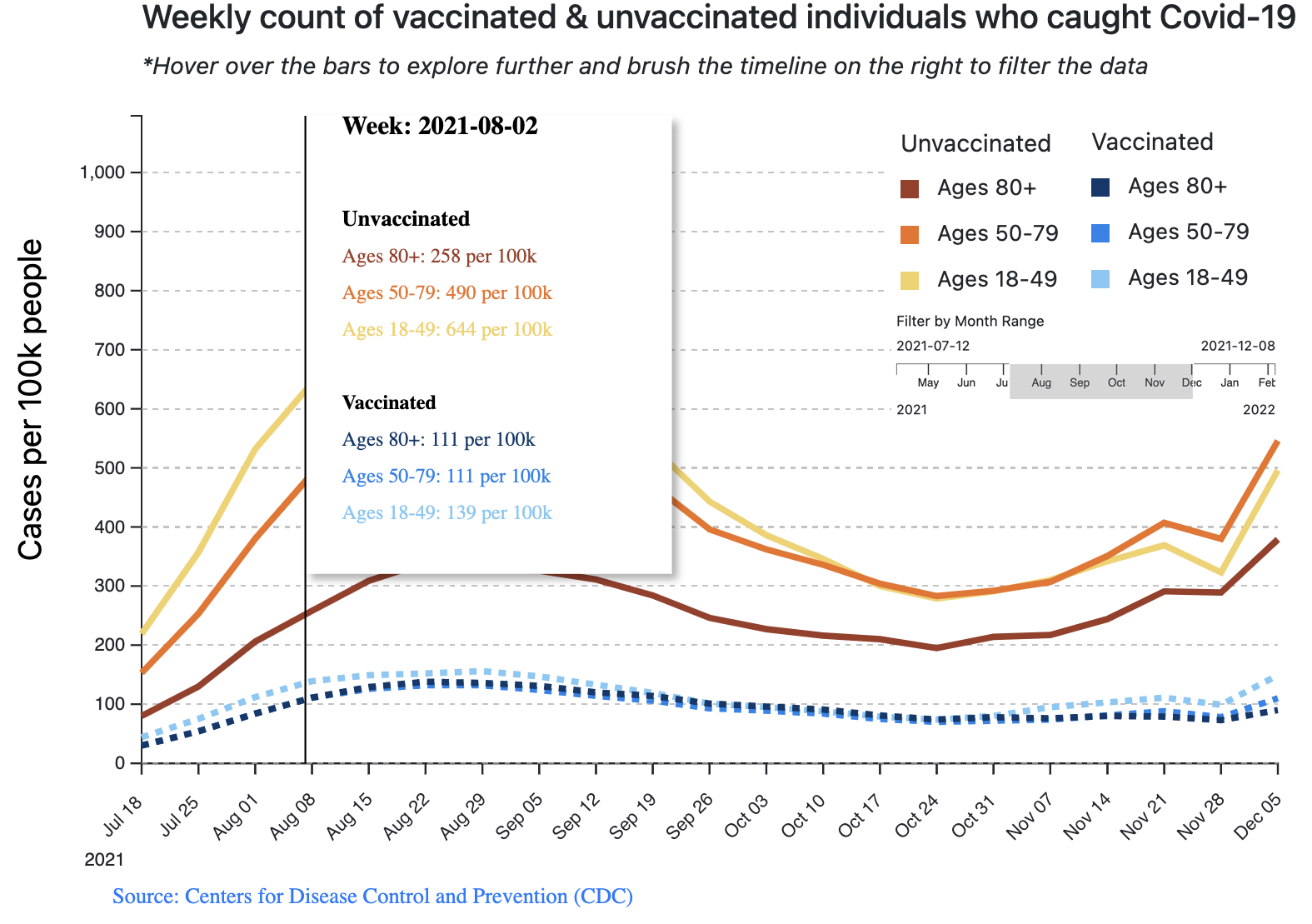}
         \caption{Line chart of high complexity.}
         \label{fig:complex_line}
     \end{subfigure}

     \vspace{-3mm}
        \caption{Study stimulus for bar charts showing Covid visualizations, across all three complexities. Figures show interactive features available to participants, namely tooltips for all three levels of visual complexity, and a time filter for the complex condition}
        \label{fig:barCharts}
             \vspace{-3mm}

\end{figure*}

\paragraph {Visual Complexity}

Our main research question for this study was whether varying levels of visual complexity affect trust in science-based visualizations. We designed three levels of complexity for each of the two chart types used in this study: simple, moderate, and complex. 

Simple visualizations show the least amount of data in the most accessible format. For bar charts, the simple visualization showed two separate bar charts, each representing one category of data (Figure \ref{fig:simple_bar}). The simple line chart contained a single chart with two lines, one for each category (Figure \ref{fig:simple_line}). Moderate visualizations provide added visual complexity, either through stacked bars - for bar charts -  (Fig \ref{fig:moderate_bar}) or by adding a confidence interval around the mean line - for line charts (Fig \ref{fig:moderate_line}). The complex condition shows the most information with 6 data categories instead of 2 for both bars (Fig \ref{fig:complex_bar}) and lines (Fig \ref{fig:complex_line}). 
% The possibility of information overload is mitigated by the added ability to interact with and filter the data to a time period of interest. 

% Add discussion on why we implemented interactivity. cite and leverage arguments made in this paper. https://ieeexplore.ieee.org/stamp/stamp.jsp?tp=&arnumber=9646526

\paragraph{Chart Types}

To assess whether our findings on visual complexity generalize to more than one chart type, we chose two of the most common visualizations used when communicating data to the general public: bar charts and line charts. We designed both visualizations to reflect standard Covid charts that have been used in the media over the past couple of years. All charts were equipped with tooltips, which showed detailed information on demand. Depending on the level of complexity, interaction was also provided via a brush filter.

% \begin{figure*}
%      \centering
%      \hspace{-3mm}
%      \begin{subfigure}[b]{0.3\textwidth}
%          % \centering
         
%          \includegraphics[height=4cm]{figs/simpleLine.png}
%          \caption{Line chart of simple complexity.}
%          \label{fig:y equals x}
%      \end{subfigure}
%      % \hfill
%      \begin{subfigure}[b]{0.3\textwidth}
%          \centering
%          \includegraphics[height=4cm]{figs/moderateLine.png}
%          \caption{Line chart of moderate complexity.}
%          \label{fig:three sin x}
%      \end{subfigure}
%      % \hfill
%      \begin{subfigure}[b]{0.3\textwidth}
%          \centering
%          \includegraphics[height=4cm]{figs/complexLine.png}
%          \caption{Line chart of high complexity.}
%          \label{fig:five over x}
%      \end{subfigure}
%         \caption{Study stimulus for line charts showing Covid visualizations, across all three complexities. Figures show interactive features available to participants, namely tooltips for all three levels of visual complexity, and a time filter for the complex condition}
%         \label{fig:lineCharts}
% \end{figure*}

\paragraph {Data Topic}

To control for prior opinions on Covid data, we created a second condition where participants were shown visualizations of Crop Diseases in Croatia. However, using two different datasets to generate the visualizations would introduce an additional confound. To avoid this, we used the same dataset for both data topics and simply changed the labels, title, and legend accordingly. Participants who were given the Croatian crops visualization saw a disclaimer at the end of the study that clarified that the data was actually for Covid cases.

\subsection{Pilots and Experiment Planning}

We conducted three pilots to evaluate tasks, interactive visualization features, measures, and our procedure. Initial pilots revealed that asking participants for their "initial impressions" of a visualization was not effective at capturing affective trust. Instead, most participants cited clearly cognitive aspects, such as trends in the data, or their takeaways. To address this, we ran a separate pre-study to establish the type of strictly affective responses participants could have to the different visualizations. We recruited 200 participants on Prolific and showed each a randomly assigned visualization from the final study for 15 seconds. This approach was based on the work done by Betella et al \cite{betella_affective_2016}, who measured human emotions toward visual stimuli by showing viewers an image for 15 seconds and then prompting them for their responses. After viewing the visualization, our participants were asked to describe the visualizations in 5 to 10 adjectives. The most common adjectives informed the dimensions of affective trust in the final study. 

We estimated the number of participants required to uncover effects based on a final pilot run on Prolific with 60 participants. We used a power analysis between the  different conditions to estimate the variance in our quantitative measures, which we combined with our observed means to estimate the number of trials required.

\vspace{-1mm}
\subsection{Participants and Procedure}

 We recruited 600 participants on Prolific, a crowdsourcing platform with a research focus. Twenty-four participants timed out and were excluded from the study. Of the remaining 576 participants, an additional 33 were excluded for either failing the attention checks or due to a glitch in the survey that did not record their response to one of the questions. This left us with 543 valid responses. Our study used a between-subjects design in which each participant was randomly assigned to one of the 12 combinations of independent variables (three visual complexities, two chart types, and two data topics). We deployed the study on the Qualtrics survey platform, which ensures random distribution amongst conditions. Using the javascript integration feature in Qualtrics allowed us to render interactive visualizations directly within the survey (as opposed to static visualizations) as well as collect interaction data from participants such as when they hovered on data elements or used the interactive filter. It also allowed us to log when they browsed away from the study, and for how long. This data was critical for quality control and to obtain more accurate time-on-task metrics for each participant. The participants were recruited using Prolific's \cite{palan_prolificacsubject_2018} representative sample feature, which ensures that participants are representative of the United States across multiple demographics, including age, sex, and ethnicity. 

 Based on completion times of pilot experiments, each participant was paid \$4 USD, for an estimated duration of 15 minutes, resulting in an hourly rate of about \$16 USD. The median time of completion after the survey was completed was 13 minutes. All participants viewed and agreed to an IRB-approved consent form. To be eligible for the study, participants had to use a laptop or desktop device with a resolution of at least 1400x850 pixels available screen space in the browser.

 The study consisted of seven sections: \textit{Affective Trust},\textit{Visualization Tour}, \textit{VLAT}, \textit{Explore the Visualization}, \textit{Trust in Data}, \textit{Trust in Visualization}, \textit{Individual Characteristics}. The full study can be viewed at \url{https://rotman.az1.qualtrics.com/jfe/form/SV_03B33TJH9p1dtTU}. 

\vspace{-2mm}

\subsection{Affective Trust}
\label{sec:affective_trust}
The first section of the study following the consent form was geared towards collecting quantitative measures of participants' affective reactions to the visualization. More specifically, we focused on quantifying \textit{affective cues towards the visualization}. Based on the results of the pre-study on affective reactions to the visualization, we asked users to move three sliders, each towards the adjective that best described the visualization (Table \ref{table:affective_sliders}). We capture other affective visualization trust components, such as \textit{benevolence} in an open text question at the end of the study. 

\begin{table}[!h]
\vspace{-2mm}
\small
\centering
\setlength{\tabcolsep}{4pt} % Default value: 6pt
\renewcommand{\arraystretch}{1.2} % Default value: 1
% \captionof{table}{Trust in Data} \label{tab:title} 
\begin{tabular}{@{}p{.1\linewidth}p{.35\linewidth}p{.4\linewidth}@{}}
% \toprule
\textbf{Type}  &  \textbf{Dimension}  & \textbf{Study Item } \\ 
\toprule
 \rowcolor{affective}
Affective & Aesthetic Cues (Scientific) & Unscientific ---------- Scientific \\
 \rowcolor{affective}
Affective & Aesthetic Cues (Clarity) & Unclear ---------------  Clear \\ 
  \rowcolor{affective}
 Affective  & Aesthetic Cues (Pretty)  & Ugly ------------------- Pretty \\

%\bottomrule
\end{tabular}
% \vspace{-3mm}
\caption{Affective Cues: List of 3 sliders given to participants to capture affective trust in the visualization through different aesthetic cues.}
\label{table:affective_sliders}
\vspace{-5mm}
\end{table}

\vspace{-1mm}
\subsection{Visualization Tour}
An interactive tour for each condition guided the user through the main components of the visualization and prompted users to interact with the visualization, including hovering or filtering. Because the brush filter in the complex visualization may be unfamiliar to novice users, we required participants to interact with the brush according to the instructions before proceeding with the tour. We also logged detailed provenance data that captured how participants interacted with the tour steps. This allowed us to perform post-hoc analysis and answer questions such as: Did they progress linearly through the tour? Did they go back to a previous step, or struggle to perform the operations with the brush filter required to progress in the study?

\subsection{VLAT}
% talk about adapting the VLAT questions to our specific visualization
To capture participants' ability to understand the visualizations, we adapted the visual literacy assessment test (VLAT) ~\cite{lee_vlat_2017} by creating equivalent questions for the visualizations used in the study.  Our intention was to create questions that required the participants to engage with the complexity of that visualization. For example, for the stacked bar charts shown in the moderate complexity condition  (see Fig. \ref{fig:moderate_bar}), we asked participants questions that required understanding the different baselines for the two data categories.  In the complex visualization, we asked a question that required participants to compare trends among different categories.  We provided multiple-choice solutions for these questions and normalized the final scores by using the formula VLAT\_Score=C-(W/N)
% \begin{equation}
%     VLAT\_Score = C - \frac{W}{N}
% \end{equation}
described in 
~\cite{lee_vlat_2017}, where $C$ is the number of questions the participant answered correctly, $W$ is the number of questions the participant answered incorrectly, and $N$ is the number of solutions in a single question. The incorrect solutions we created were intended to be intuitive or nearly correct. 

% At this stage of the study, participants were asked to answer the VLAT questions for their assigned complexity condition. After they completed the trust in data (Sec. \ref{sec:trust_in_data}) and visualization (Sec. \ref{sec:trust_in_vis} sections, participants were asked to complete the VLAT for the other two complexity conditions. This data allows us to consider their general visual literacy as well as the specific literacy for their complexity condition.

\vspace{-1mm}

\subsection {Explore the Visualization}
Participant interaction data logged during one of the pilot studies revealed that participants were not spending any time engaging with or analyzing the visualization before reporting on their trust levels. Since cognitive trust antecedents rely on processing the visualization at a non-superficial level, we added this ``explore section", which asked participants to look at/interact with the visualization for 15 seconds before allowing them to proceed to the sections on trust assessments. During this time, we continued to collect interaction provenance data including the data of visual elements that they hovered over and for how long, the start and end points of the brush filters they created, and whether they browse away from the study. Each of these provenance data was stored with a timestamp for when they occurred.

\vspace{-2mm}
\subsection{Trust in Visualization}
\label{sec:trust_in_vis}
% Point out the mapping between the framework concepts and the items in the this quetsion
In the trust in visualization section participants were asked to \textit{assume they trust the data}, and rate their trust in the visualization. Trust in the visualization was captured via a 6-item set of Likert scales (Table \ref{table:vis_items}). The first three items capture cognitive and affective trust antecedents for trust in the visualization. We describe the fourth and fifth items in Section \ref{sec:behavior}.

\begin{table}[!h]
\vspace{-2mm}
\small
\centering
\setlength{\tabcolsep}{4pt} % Default value: 6pt
\renewcommand{\arraystretch}{1.2} % Default value: 1
% \captionof{table}{Trust in Data} \label{tab:title} 
\begin{tabular}{@{}p{.1\linewidth}p{.2\linewidth}p{.6\linewidth}@{}}
% \toprule
\textbf{Type}  &  \textbf{Dimension}  & \textbf{Study Item } \\ 
\toprule
\rowcolor{cognitive}
 Cognitive & Accuracy & The visualization transparently includes all important elements of the data \\ 
 \rowcolor{cognitive}
Cognitive & Clarity & I find it easy to understand this visualization\\ 
 \rowcolor{affective}
Affective & Aesthetic Cues & I like this visualization \\ 
 \rowcolor{behavior}
 Behavior & Actionable & I would likely use this visualization and its information in my daily life \\
 \rowcolor{behavior}
 Behavior & Shareable & I would likely share this visualization with my family, friends or on social media\\ 
 \hline
  \rowcolor{trustColor}
  Trust & Visualization  & I trust this visualization \\ 
\hline

%\bottomrule
\end{tabular}
% \vspace{-3mm}
\caption{Trust in Visualization: List of 6 Likert scale items given to participants to capture trust in visualization. Items 1-3 capture cognitive and affective antecedents, and Items 4-5 capture behavioral outcomes. Item 6 captures the high-level trust in the visualization.}
\label{table:vis_items}
\vspace{-3mm}
\end{table}

% Two items captured cognitive trust, which included their assessment of the visualization accuracy, and clarity. One item captured affective trust. Two items assess whether participants would be willing to act according to the information displayed by the visualization. These statements aim to capture the 'behavior' component of the trust framework, which is an outcome of trusting information that is given to you.  

The first five items of this section were displayed in random order to mitigate the effect of the order on response, and the final item, ``I trust this visualization," is shown last regardless of the participant, as we hypothesize that general feelings of trust in the visualization are a combination of the beliefs expressed by the previous five items.

At the end of the trust in visualization section, participants were asked to provide a text explanation of their trust rating. This qualitative data allowed us to capture both cognitive and affective trust dimensions beyond those in the 6-item Likert scales. This includes elements such as usability, benevolence, ethics, and affective cues beyond the ones captured in Section \ref{sec:affective_trust}.

Participants saw this and the next section (Trust in Data) in random order to mitigate any priming that may come from the ordering.

\subsection{Trust in Data}
\label{sec:trust_in_data}

In this section, the participants were asked to disregard the visualization and rate their trust in the underlying data. As with the trust in visualization section, trust in data was captured via a 6-item set of Likert scales (Table \ref{table:data_items}).  We did not explicitly elicit ratings of data currency, since the data time frame was fixed from Dec 2020 to Dec 2021 for this study.

The first five items were again displayed in random order to mitigate the effect of the order on response. The final item, ``I trust this data," is always shown last, as we similarly hypothesize that general feelings of trust in the data are a combination of the beliefs expressed by the previous five items.

\begin{table}[!h]
\vspace{-2mm}
\small
\centering
\setlength{\tabcolsep}{4pt} % Default value: 6pt
\renewcommand{\arraystretch}{1.2} % Default value: 1
% \captionof{table}{Trust in Data} \label{tab:title} 
\begin{tabular}{@{}p{.1\linewidth}p{.2\linewidth}p{.6\linewidth}@{}}
% \toprule
\textbf{Type}  &  \textbf{Dimension}  & \textbf{Study Item } \\ 
\toprule
\rowcolor{cognitive}
 Cognitive & Accuracy & The data is accurate \\ 
 \rowcolor{cognitive}
Cognitive & Coverage & The data is complete and does not leave out important information \\ 
 \rowcolor{cognitive}
Cognitive & Clarity & I understand the meaning of this data well \\ 
 \rowcolor{affective}
Affective & Benevolence & The data is unbiased and trustworthy \\
 \rowcolor{affective}
Affective & Affective Cues & The data source was clearly displayed\\ 
\hline
  \rowcolor{trustColor}
 Trust  & Data  & I trust this data \\ 
 \hline

%\bottomrule
\end{tabular}
% \vspace{-3mm}
\caption{Trust in Data: List of 6 Likert scale items given to participants to capture trust in the data. Items 1-5 capture cognitive and affective antecedents to trust in data. Item 6 captures the high-level trust in the data.}
\label{table:data_items}
\vspace{-5mm}
\end{table}

\subsection {Individual Characteristics}

As discussed in our trust framework (Section \ref{sec:trust_framework}), individual differences can greatly impact trust formation. In this study, we operationalize these differences by capturing a broad set of participant traits. When analyzing the study results, we include individual characteristics as co-variates in our regression models. This allows us to better isolate the magnitude of trust variations that are due to changes in our controlled variables, and not due to individual characteristics of the participants. 

\paragraph{Interpersonal Trust}
% One of the individual characteristics 

A person's generalized baseline trust in people can be used to compare their ratings of trust in the specific context of data and visualizations. Hence, we use the ITS interpersonal trust scale \cite{evans_survey_2008} and asked participants to answer the question ``Generally speaking, would you say that most people can be trusted or that you can't be too careful in dealing with people?" on a scale from 1=``Most people cannot be trusted" to 7=``Most people can be trusted."

\paragraph{Trust in Institutions}
% One of the individual characteristics 

Beliefs regarding Covid-19 are often impacted by an individual's trust in related institutions (e.g., government, science, the new media, etc). Hence, we asked the participants to rate their trust in the following institutions on a scale from 0 to 10: (1) political parties, (2) the government, (3) the police, (4) the legal system, (5) the new media, (6) business and industry, (7) scientists/science, and (8) doctors.

\paragraph{Need for Cognition}
% One of the individual characteristics 

Participant responses to the cognitive trust items of our framework may be influenced by their motivation to engage in cognitive activity in general (e.g., someone who is less motivated to engage in cognitive activity may rate the cognitive trust items lower). Thus, we employ the 6-item version of the Need for Cognition test, as proposed by Coelho et. al \cite{lins_de_holanda_coelho_very_2018}. For this participants rated how strongly they agreed/disagreed with statements such as ``I would prefer complex to simple problems."

\paragraph{Politics}
% One of the individual characteristics 

Because one of the datasets we use for the visualization in this study is related to Covid-19, a heavily politicized topic, we collect information regarding the political affiliation (7-point scale from ``extremely liberal" to ``extremely conservative")  and Covid-19 media consumption of the participants, i.e., how often they sought out information on Covid during the pandemic). For the participants that did not see the Covid-19 visualization, we did not ask about their Covid-19 media habits.

\paragraph{Demographics}
% One of the individual characteristics 

To account for any effect of participant demographics on their trust behavior, we collect information regarding participants' age, gender, state residency, education level, parents' education level, spoken language, ethnicity, income, and religious affiliation.

\subsection{Behavioral Outcomes}
\label{sec:behavior}

We captured participants' self-reported behavioral outcomes (Figure \ref{fig:teaser}) by asking them to rate the items ``I would likely use this visualization and its information in my daily life" and ``I would likely share this visualization with my family, friends or on social media" (Table \ref{table:vis_items}). While self-reported measures are not always reliable indicators of actual behavioral change \cite{dang_why_2020}, capturing intent provided us with initial insight into ways in which trust might impact behavior. In Section \ref{sec:predicting_behavior}, we analyze how trust in the visualization and underlying data correlated with participant responses to these items.

\section {Study Results}
\label{sec:results}

The data collected in this study includes both individual characteristics of each participant -- such as visual literacy, need for cognition, and demographic information -- as well as the perceived cognitive and affective trust of the visualizations. We analyzed this data by constructing linear regression models between properties in our framework. 
The independent variables in our models are chart complexity, data topic, and chart type.
We also include demographic information (e.g., gender, education, ethnicity, etc), need for cognition, interpersonal trust, and trust in science as covariates to account for individual differences.
In models where we are interested in the predictive power of the trust antecedents on overall trust, we include the trust antecedents as independent variables. 
Below, we report on our major findings, which reflect the linear models for all 12 conditions (3 visual complexities, 2 chart types, and 2 data topics).  The complete dataset collected in the study, the analysis scripts, and the code used to generate the interactive visualizations are open-source and available at \url{"https://osf.io/frbaj/?view_only=de44feae51374a71ac2a9b8798559d24"}. 
% We contend that we have not exhausted the possible analysis of this data and suggest future work to leverage this dataset to further our understanding of trust mechanisms in data visualization.

%%%%%%%%%%%%%%%%%%%%%%%%%%%%%%%%%%%%%%%%%%%%%%%%%%%%%%%%%%%%%%%%%%%
% \vspace{-2mm}
\subsection{Predictive Power of Trust Antecedents}
\label{sec:antecedents}

% need to include overall trust item ratings summary stats here

As discussed in Sections \ref{sec:trust_in_vis} and \ref{sec:trust_in_data}, the study was designed to capture high-level trust in vis and data as well as the different antecedents and behavioral outcomes of trust outlined in the trust framework. We measured these aspects via the scale items shown in Tables \ref{table:affective_sliders},  \ref{table:vis_items}, and \ref{table:data_items}. 

% For trust in the visualization, items 1-3 capture cognitive and affective antecedents and items 4-5 capture behavioral outcomes. Item 6 captures the high level trust in the visualization. For trust in the data, items 1-5 capture cognitive and affective antecedents to trust in data. Item 6 captures the high level trust in the data

\subsubsection*{Which Antecedents Predict Trust? }
To examine the predictive power of the antecedents on overall trust in the visualizations and their underlying data, we constructed a linear regression model predicting visualization trust with all vis antecedents, and another model predicting data trust with all data antecedents. Demographic and individual characteristic measures were included to account for their covariances. The results of these models are shown in Table \ref{table:antecedents_predictive_power}.

\begin{table}[!h]
\vspace{-2mm}
\small
\centering
\setlength{\tabcolsep}{4pt} % Default value: 6pt
\renewcommand{\arraystretch}{1.2} % Default value: 1
% \captionof{table}{Trust in Data} \label{tab:title} 
\begin{tabular}{@{}p{.15\linewidth}p{.3\linewidth}p{.15\linewidth}p{.15\linewidth}p{.1\linewidth}@{}}
% \toprule
% \rowcolor{trustColor}
  \multicolumn{2}{l|}{\textbf{Visualization Trust Antecedent}}& \multicolumn{3}{l}{\textbf{Predictive Power on Vis Trust}} \\ 
 
\hline
\textbf{Type}  &  \textbf{Dimension}  & \textbf{Est} & \textbf{SE} & \textbf{P} \\ 
\toprule
\rowcolor{cognitive}
Cognitive  & Accuracy & 2.361e-01 & 3.758e-02 & \textcolor{significant}{\textbf{8.02e-10}}  \\ 
 \rowcolor{cognitive}
Cognitive  &  Clarity & 1.419e-01 & 4.806e-02 &  \textcolor{significant}{\textbf{0.00333}}  \\ 
\rowcolor{affective}
 Affective  &  Aesthetic Cues (Like)  & 2.292e-01 & 4.631e-02 &  \textcolor{significant}{ \textbf{1.07e-06}} \\
 \rowcolor{affective}
 Affective  &  Aesthetic Cues (Scientific) & 6.597e-03  & 2.573e-03 &  \textcolor{significant}{ \textbf{0.01067}} \\
 \rowcolor{affective}
 Affective  &  Aesthetic Cues  (Clarity) & -3.865e-04 & 1.966e-03 &  0.84425  \\
 \rowcolor{affective}
 Affective  &  Aesthetic Cues (Pretty) & -1.326e-03 & 2.047e-03 & 0.51744 \\

% UPDATED RESULTS WITH CORRECTED DATA 
% TABLE COPIED OVER BY CAROLINA ON JULY 21 
%lm(formula = vis.trust_6 ~ vis.trust_1 + vis.trust_2 + vis.trust_3 + 
%     vis.trust_4 + vis.trust_5 + affect.science_1 + affect.clarity_1 + 
%     affect.aesthetic_1 + Age + Gender + State_1 + Education + 
%     Parents_education + Language + Ethnicity + Income + Religion + 
%     trust.in.science_7 + need_for_cognition + interpersonal.trust_1, 
%     data = results)

% Residuals:
%      Min       1Q   Median       3Q      Max 
% -2.83565 -0.50664  0.01138  0.51001  2.58176 

% Coefficients: (1 not defined because of singularities)
%                         Estimate Std. Error t value Pr(>|t|)    
% (Intercept)           -11.568583   5.984157  -1.933 0.053853 .  
% vis.trust_1             0.215786   0.038071   5.668 2.62e-08 ***
% vis.trust_2             0.150292   0.047905   3.137 0.001820 ** 
% vis.trust_3             0.188862   0.048434   3.899 0.000111 ***
% vis.trust_4             0.056337   0.034117   1.651 0.099392 .  
% vis.trust_5             0.020632   0.033502   0.616 0.538318    
% affect.science_1        0.006861   0.002558   2.682 0.007601 ** 
% affect.clarity_1       -0.000636   0.001958  -0.325 0.745437    
% affect.aesthetic_1     -0.001043   0.002041  -0.511 0.609684    

 \multicolumn{4}{l}{}\\
% \rowcolor{trustColor}
  \multicolumn{2}{l|}{\textbf{Data Trust Antecedent}}& \multicolumn{3}{l}{\textbf{Predictive Power on Data Trust}} \\ 
 
\hline
 \textbf{Type}  &  \textbf{Dimension}  & \textbf{Est} & \textbf{SE} & \textbf{P} \\ \toprule
\rowcolor{cognitive}
Cognitive  &  Accuracy & 5.598e-01 & 4.243e-02 &  \textcolor{significant}{\textbf{2e-16}}   \\ 
 \rowcolor{cognitive}
 Cognitive  &  Coverage & 1.124e-01 & 3.145e-02 &\textcolor{significant}{ \textbf{0.000392}}  \\
\rowcolor{cognitive}
 Cognitive  &  Clarity & 1.875e-01 & 3.387e-02  &  \textcolor{significant}{\textbf{5.33e-08}} \\
\rowcolor{affective}
  Affective  &   Benevolence & 1.912e-02 & 3.519e-02  & 0.587117 \\
  \rowcolor{affective}
  Affective  &   Affective Cues (Source) & 6.567e-02   & 3.099e-02 & \textcolor{significant}{\textbf{0.034662}} \\

% UPDATED RESULTS WITH CORRECTED DATA 
% TABLE COPIED OVER BY CAROLINA ON JULY 21 
% lm(formula = data.trust_6 ~ data.trust_1 + data.trust_2 + data.trust_3 + 
%     data.trust_4 + data.trust_5 + Age + Gender + State_1 + Education + 
%     Parents_education + Language + Ethnicity + Income + Religion + 
%     trust.in.science_7 + need_for_cognition + interpersonal.trust_1, 
%     data = results)

% Residuals:
%     Min      1Q  Median      3Q     Max 
% -2.2352 -0.3538  0.0000  0.4104  2.2492 

% Coefficients: (1 not defined because of singularities)
%                         Estimate Std. Error t value Pr(>|t|)    
% (Intercept)           -1.316e+01  5.014e+00  -2.623 0.009005 ** 
% data.trust_1           5.598e-01  4.243e-02  13.194  < 2e-16 ***
% data.trust_2           1.124e-01  3.145e-02   3.573 0.000392 ***
% data.trust_3           1.875e-01  3.387e-02   5.535 5.33e-08 ***
% data.trust_4           1.912e-02  3.519e-02   0.543 0.587117    
% data.trust_5           6.567e-02  3.099e-02   2.119 0.034662 *  

%\bottomrule
\end{tabular}
% \vspace{-3mm}
\caption{Results of linear regressions modeling the predictive power of trust antecedents in predicting trust in the visualization and trust in data. The columns refer to the following: Est is the estimated slope of the linear regression, SE is the standard error, and P is the p-value. Significant p-values are highlighted in red.}
\label{table:antecedents_predictive_power}
\vspace{-3mm}
\end{table}

%\pheading{Trust in Vis:} 
\textbf{Trust in Visualization}
From our analysis of the antecedents, we find that two cognitive trust antecedents and two affective trust antecedents are significant predictors of trust in visualization. 
\textit{Cognitive accuracy} and \textit{cognitive clarity} were significantly correlated with the overall trust in the visualization. 
Similarly, responses to the \textit{aesthetic cues} and the bipolar scale from ``Unscientific" to ``Scientific" were significantly correlated with overall trust in the visualization. Generally, \textbf{participants were likely to trust the visualization if they had found it scientific, accurate, and easy to understand, as well as if they had a positive affective experience}.

\textbf{Trust in Data}
We also analyzed the predictive power of cognitive and affective trust antecedents on trust in the underlying data and found that \textit{cognitive accuracy}, \textit{cognitive clarity}, \textit{cognitive coverage} and \textit{data source} were significantly correlated with overall trust in the data. Hence, \textbf{participants were more likely to trust the underlying data from a visualization if they found it easy to understand, and complete, and if the source of the data was clearly displayed}.

\subsubsection*{Which Trust Measurements Predict Behavioral Outcomes?}
\label{sec:predicting_behavior}

Though capturing behavioral outcomes as a result of trust in visualization would be best achieved with a longitudinal study, our study includes two items that measure projected behavioral outcomes through self-reported measures: (1) How likely it is the participant would use the visualization in their daily life, and (2) How likely it is the participant would share the visualization with friends and family. 

We investigate which of the trust measurements, whether antecedents or high-level trust, serve as predictors for these behavioral metrics. For example, are participants more likely to use or share the visualization if they find the visualization clear? Or if they find the data unbiased? 

We found that participants who highly rated the \textit{cognitive accuracy} and \textit{aesthetic cues} of the visualization were significantly more likely to both use the visualization in their daily life and share it with family and friends. Additionally, participants who rated the visualization as more clear on the bipolar slider from ``Unclear" to ``Clear" (\textit{affective clarity}) were significantly more likely to use the visualization in their daily life.

%%%%%%%%%%%%%%%%%%%%%%%%%%%%%%%%%%%%%%%%%%%%%%%%%%%%%%%%%%%%%%%%%%%
\subsection{Correlation and VIF between Trust Items}
\label{sec:vif}

We used the variance inflation factor to analyze the multicollinearity between the eleven antecedents to trust, the two measures of overall trust, interpersonal trust, need for cognition and trust in science. The complete matrix of VIF scores between all variable pairs is included in the supplementary material. Although the metrics are not fully independent, all VIFs, except for one, are below 3.4, which indicates a low level of colinearity \cite{kock_lateral_2012}. Trust in data had a slightly higher VIF of 4.3, suggesting that participants' trust in the data can overlap with trust in the visualization, which is expected given the fact the visualization is meant to represent the underlying data. 
% except for overall trust in visualization which is 2.342. There is some debate about the accepted threshold for colinearity; commonly accepted thresholds are 10, 5, and 3.3 \cite{kock2012lateral}. Even using the lowest threshold commonly recommended, all of our VIFs are below the threshold. 
Trust in visualization has a VIF of 3.16, which suggests that participants' overall trust in the visualization overlaps with both visualization and data trust antecedents, as well as the three personality traits. 
This potentially suggests that overall trust in the visualization is impacted by trust in data, and is relatively well-captured by the eleven antecedents.
Importantly, the personality traits (interpersonal trust, need for cognition, and trust in science) have very low VIFs (close to 1.0), indicating a low overlap with the other metrics considered. 

% is more influenced by trust in the underlying data, and trust in data is less impacted by trust in visualizations.

%%%%%%%%%%%%%%%%%%%%%%%%%%%%%%%%%%%%%%%%%%%%%%%%%%%%%%%%%%%%%%%%%%%
\subsection{Trust in Visualization and Trust in Data}

% trust across chart types and data topics
% Trust and Complexity + ChartType + isCovidData

\subsubsection*{Does Visual Complexity Influence Trust in Visualization?}

% Our intial hypothesis, when designing the study, was that changes in the complexity of the visualization would impact viewer's trust. 
As described in Section \ref{sec:study_design}, we designed two types of visualizations (bar charts and line charts) in three levels of complexity: simple, moderate, or complex. In this section, we examine the effect that different levels of visual complexity have on perceived trust in the visualization. To test the generalizability of our effect, we investigate whether the effect of chart complexity varies across the two chart types (bar, line) and two data topics (Covid data, crop disease data).

\begin{table}[!h]
\vspace{-2mm}
\small
\centering
\setlength{\tabcolsep}{4pt} % Default value: 6pt
\renewcommand{\arraystretch}{1.2} % Default value: 1
% \captionof{table}{Trust in Data} \label{tab:title} 
\begin{tabular}{@{}p{.235\linewidth}p{.2\linewidth}p{.1\linewidth}p{.2\linewidth}p{.12\linewidth}@{}}
% \toprule
\rowcolor{trustColor}
 & \multicolumn{2}{l|}{\textbf{a. trust in the visualization}} & \multicolumn{2}{l}{\textbf{b. trust in the data}} \\ 

\hline
\textbf{Predictor}   & \textbf{F Value} & \multicolumn{1}{l|}{\textbf{Pr(>F)}} 
& \textbf{F Value }  & \textbf{Pr(>F)} \\ 
\toprule
%                                     Df  F value      Pr(>F) 
% complexity                           2   4.0798   2.919e-08    
chart complexity     & F(2,437)=4.0798   & \multicolumn{1}{l|}{\textcolor{significant}{\textbf{2.919e-08}} }
%                                      2   2.5088   0.003029
                     & F(2,437)=2.5088   & \textcolor{significant}{\textbf{0.003029}} \\ 
% as.factor(isCovidData)               1   8.3574   1.621e-11
data topic            & F(1,437)=8.3574  & \multicolumn{1}{l|}{\textcolor{significant}{\textbf{1.621e-11}}}
%                                      1  10.5293   6.446e-11
                     & F(1,437)=10.9253   &\textcolor{significant}{\textbf{6.446e-11}} \\ 
% chartType                            1   1.5884   0.116163 
chart type           & F(1,437)=1.5884  & \multicolumn{1}{l|}{0.116163} 
%                                      1   1.1763   0.317789
                     & F(1,437)=1.1763   & {0.317789} \\ 
% complexity*chartType                 2   1.7620   0.025638
complexity*chartType & F(1,437)=1.7620  & \multicolumn{1}{l|}{\textcolor{significant}{\textbf{0.025638}}} 
%                                      2   1.2231   0.261815
                     & F(1,437)=1.2231   & {0.261815}  \\ \\ 
% Age                                  1   1.9079   0.049143
Age                  & F(1,437)=1.9079   & \multicolumn{1}{l|}{\textcolor{significant}{\textbf{0.049143}}}
%                                      1   2.6696   0.014868
                     & F(1,437)=2.6696   & \textcolor{significant}{\textbf{0.014868}} \\
% Gender                               3   0.9113   0.596349
Gender               & F(3,437)=0.9113   & \multicolumn{1}{l|}{0.596349}
%                                      3   1.6250   0.047127
                     & F(3,437)=1.6250   & \textcolor{significant}{\textbf{0.047127}} \\
% State_1                             46   1.1513   0.023586
State                & F(46,437)=1.1513  & \multicolumn{1}{l|}{\textcolor{significant}{\textbf{0.023586}}}
%                                     46   1.2542   0.004228
                     & F(46,437)=1.2542  & \textcolor{significant}{\textbf{0.004228}} \\
% as.factor(Education)                 9   1.2956   0.039962
Education            & F(9,437)=1.2956   & \multicolumn{1}{l|}{\textcolor{significant}{\textbf{0.039962}}}
%                                      9   1.2516   0.104194
                     & F(9,437)=1.2516   & 0.104194 \\
% trust.in.science_7                   1  13.1355  < 2.2e-16
Trust in Science     & F(1,437)=13.1355  & \multicolumn{1}{l|}{\textcolor{significant}{\textbf{< 2.2e-16}}} 
%                                      1  19.2062  < 2.2e-16
                     & F(1,437)=19.2062  & \textcolor{significant}{\textbf{< 2.2e-16}}\\
% need_for_cognition                   1   2.8732   0.002644
Need for Cognition   & F(1,437)=2.8732   & \multicolumn{1}{l|}{\textcolor{significant}{\textbf{0.002644}}} 
%                                      1   3.2527   0.003876
                     & F(1,437)=3.2527   & \textcolor{significant}{\textbf{0.003876}} \\
% interpersonal.trust_1                1   2.1778   0.022599
Interpersonal Trust  & F(1,437)=2.1778   & \multicolumn{1}{l|}{\textcolor{significant}{\textbf{0.022599}}}
%                                      1   2.3003   0.033816
                     & F(1,437)=2.3003   & \textcolor{significant}{\textbf{0.033816}} \\

%\bottomrule
\end{tabular}
% \vspace{-3mm}
\caption{Results of linear regression models predicting trust in the visualization/trust in the data, with chart complexity, data topic, chart type, and demographic/individual characteristics. The column names refer to the following: F Value refers to the effect size, Pr(>F) refers to the p-value. Significant p-values are highlighted in red.}
\label{table:trust_in_vis_data_stats}
\vspace{-3mm}
\end{table}

%Response: vis.trust_6
%  ** 
%     
% Residuals                                   436 523.04   1.200 

We constructed a linear regression model predicting trust in the visualization (as measured by the 6 items shown in Table \ref{table:vis_items}) with chart complexity, data topic, chart type, and their interactions, including the three-way interaction (Table \ref{table:trust_in_vis_data_stats}). We also added demographic and characteristic predictors (participants' age, gender, state of residence, education level, their parent's education level, fluent language, ethnicity, income, religion, level of trust in science, general need for cognition, tendency for interpersonal trust) as co-variates to the model to account for their impact. We report the effect size and p-value for all model predictors in Table \ref{table:trust_in_vis_data_stats}.

Our results show that ``Trust in science" has the largest effect size in the model for both trust in visualization and trust in data. This is somewhat expected since both visualizations were designed to communicate science to the general public. Including the effect of trust in science in our models allowed us to determine the smaller, yet still significant, roles of our independent variables (e.g., visual complexity and data topic). The VIF analysis described in Section \ref{sec:vif} also shows low collinearity between trust in science and the other variables in the model, further indicating that while trust in science and the underlying institutions play a significant role in establishing trust, the visual components studied can still modulate trust formation in a significant way. 

% In analyzing the trust ratings of participants who saw the COVID data (N = 272), we observed a significant main effect of complexity on trust ratings (INSERT STATS), such that participants who saw the simple chart gave higher trust ratings, as shown in Figure~\ref{fig:vis_complexityDataChart}
% This main effect goes away when we looked at the trust ratings of a different set of participants who saw the same data labeled to be about crop diseases in Croatia (N = 277, INSERT STATS).
% To compare the sizes of the two main effects, we created an interaction model considering data type (COVID or crop) and complexity, combining participants from both experiments while keeping all the other co-variates constant in the model. 
% The results suggest that while complexity can predict trust in the COVID data condition, its effect size is overpowered by the between-subject variance created by the topic variable from the two pools of participants, as the interaction is not significant (STATS). 

\begin{figure}[!h]
\vspace{-3mm}

    % \centering
    \includegraphics[width = \linewidth]{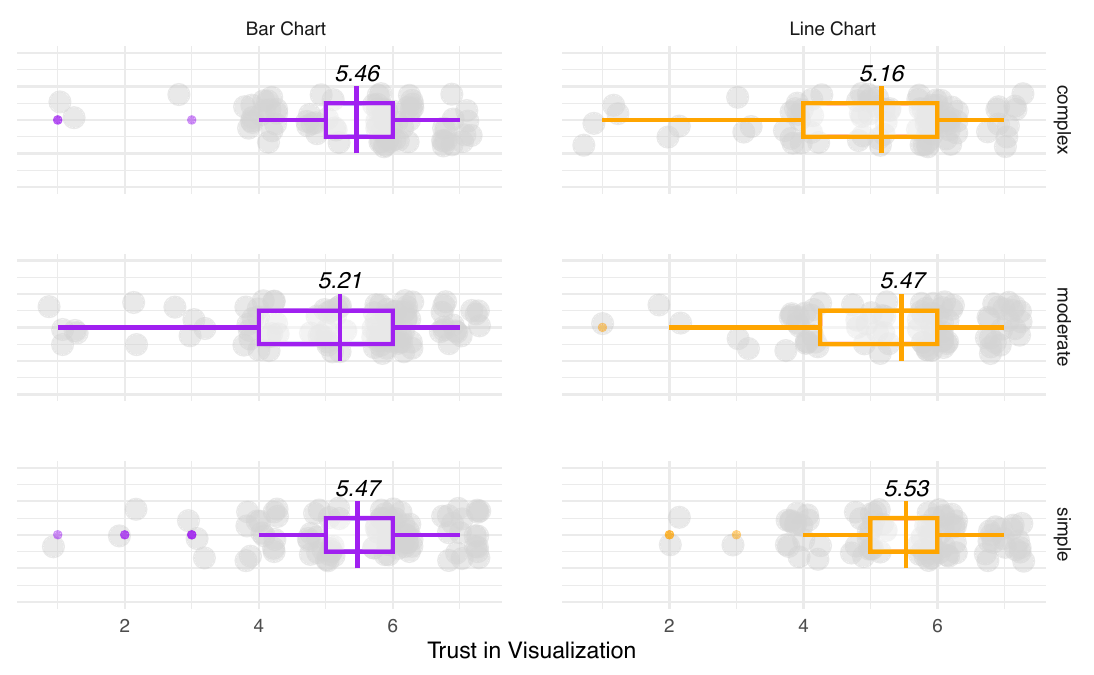}
    \caption{Trust in the visualization as a function of chart type and visual complexity. The visualization shows a box and whisker plot. The central line shows the mean of the distribution with a corresponding text label. The whiskers show the quartiles and extrema. The grey dots visualize all of the data points.}
    \label{fig:vis_complexityDataChart}
    % \vspace{-5mm}
\end{figure}

 Our results show \textbf{a main effect of visual complexity and data topic on participants' trust ratings for the visualization.} as shown in Table \ref{table:trust_in_vis_data_stats} (column a). We ran independent linear models of each of the 6 trust measurements (see complete tables in the supplementary materials)to identify which trust antecedents drive the main effects observed in the overall trust model. The results show that the complexity effect is driven by participants' response to the antecedents related to clarity, both cognitive and affective. Specifically, participants found the simple visualizations significantly easier to understand than the more complex visualizations. As discussed in Section \ref{sec:antecedents}, clarity of the visualization is a significant predictor of overall trust.\textbf{ That is, participants were more likely to trust visualizations that they found clear and easy to understand.} The effect of the data topic on trust is driven by participants' responses to self-reported behavioral intentions. Specifically, participants were more likely to share and use visualizations related to a topic that mattered to them (COVID) than one that did not (Croatian Crops). We also observed \textbf{a significant interaction between chart complexity and chart type}.  This interaction was driven by participants' responses to the affective antecedent on how much they liked each chart type.  \textbf{Specifically, participants liked line charts more than bar charts and reported decreasing trust ratings as the line charts became more complex} (Figure~\ref{fig:vis_complexityDataChart}). The complexity effect was not observed for participants who saw the bar charts.

% As described in our proposed trust framework and evidenced by our analysis in Section \ref{sec:antecedents}, trust is driven by a series of both cognitive antecedents (e.g., visualization accuracy) and affective antecedents (e.g., perceived benevolence of the visualization creator).
% To determine which trust antecedents were a driving force of the interaction effect of data topic and complexity on trust, we constructed a linear regression model for each antecedent of trust in visualization with a predictor for chart complexity. We filtered the data for this model to only consider participants who saw Covid visualizations since they exhibited a significant effect of complexity on overall trust in visualizations. Detailed results are available in the supplementary material. We discuss these results further in Section \ref{sec:discussion_case_study}.

\subsubsection*{Does Visual Complexity Influence Trust in the Underlying Data?}

\begin{figure}[!h]
% \vspace{-3mm}

    % \centering
    \includegraphics[width = \linewidth]{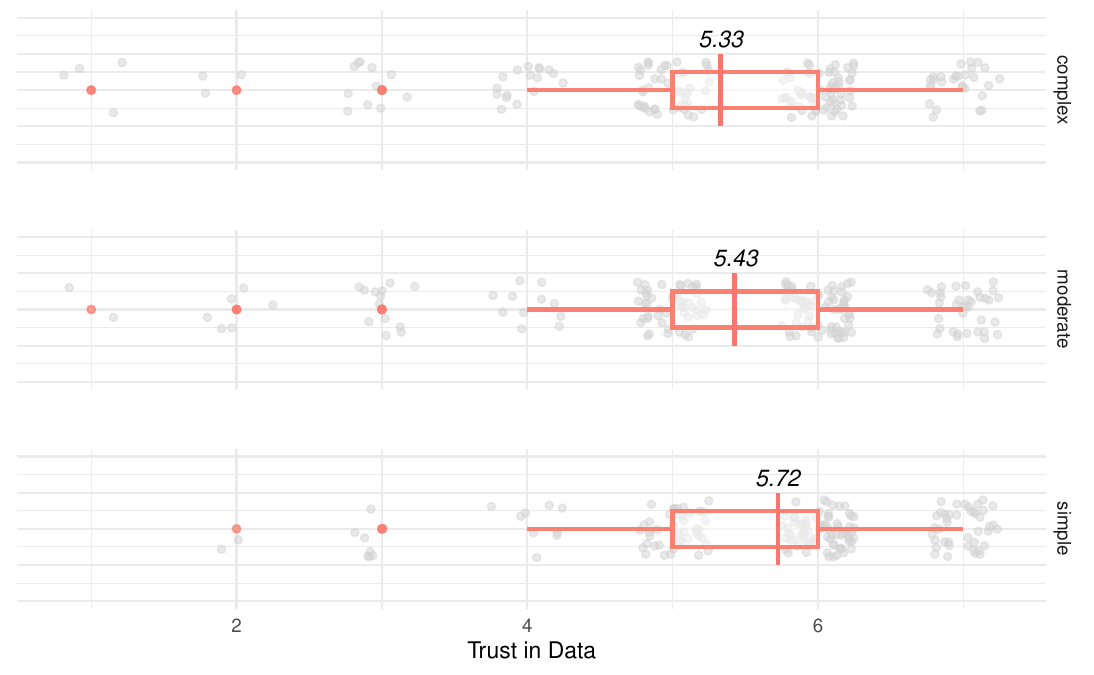}
    \caption{Trust in the data as a function of visual complexity. The visualization shows a box and whisker plot. The central line shows the mean of the distribution with a corresponding text label. The whiskers show the quartiles and extrema. The grey dots visualize all of the data points.}
    \label{fig:data_complexityDataChart}
    % \vspace{-3mm}
\end{figure}

We constructed a similar linear regression model predicting trust in the data. We used the same demographic and characteristic predictors as co-variate factors. The effect sizes and p values for all predictors are reported in Table \ref{table:trust_in_vis_data_stats} (column b). Similar to the trust in visualization analysis, we found\textbf{ a main effect of chart complexity and data topic on trust in the data }. Independent linear models of each of the 6 data trust measurements show that the complexity effect is driven by participants' response to the affective metrics on benevolence (``The data is unbiased and trustworthy") and source (``The data source was clearly displayed"). The effect of complexity on the source antecedent was surprising since the data source was shown equally for all visualizations. We speculate that participants were less likely to spot the source of the data in complex visualizations, therefore rating the source transparency lower in those cases.  In regards to the data benevolence, participants gave simple visualizations the highest ratings, and as the visualizations became more complex, ratings for data trustworthiness decreased. (Figure \ref{fig:data_complexityDataChart}). \textbf{ That is, participants were more likely to trust the data for simple visualizations than complex ones.}

The effect of data topic on the trust in data was also driven by participants' response to the affective metrics on benevolence. Specifically, participants who saw the COVID data topic rated the data significantly more trustworthy than those who saw the data on Croatian Crops. Analysis of participants' qualitative responses to their trust ratings suggests that \textbf{participants were more likely to trust the data for a topic they were familiar with than an unfamiliar one}.

\section {Discussion of Case Study Results}
\label{sec:discussion_case_study}
The study data provided rich insight into how trust is modulated by visual complexity. We emphasize three main takeaways from our quantitative analysis. The first two relate to how trust in the visualization can be modulated by visual complexity and different chart types. The third takeaway relates to the role of complexity in establishing trust in the data.  As described in Section \ref{sec:results}, in our results for both trust in visualization and trust in data, we also observed a main effect of the data topic (COVID-19 and the Croatian crop data). We designed the study to control for differences in the relevance of data topics. However, differences in participant trust between these two conditions could also be a result of trust in the underlying data source or institution creating the visualization. Future studies could explore the effect of different data sources/visualization creators compared to data topics when forming trust judgments. Along these lines, existing work on trust in COVID-19 data has shown that different data sources can impact trust formation depending on the demographics and region \cite{zhang_shifting_2022}.

The first takeaway is that \textbf{participants were more likely to trust visualizations that they found clear and easy to understand}. This aspect of trust was captured in two of the trust in visualization antecedents, one cognitive (``I find it easy to understand this visualization") and one affective (Participants were asked to rank how clear they found the visualization after seeing it for the first time and for only 15 seconds). This finding is further supported by qualitative responses such as \textit{``The visualization is easy to understand and gather info from, and I trust that the information provided is accurate. "} and \textit{``I think it's really easy to follow and it appears to be realistic information that could potentially happen."}

% In our study, participants that saw the visualizations on crop diseases in Croatia were not influenced by the complexity of the visualization when determining their level of trust.
% Conversely, participants who saw Covid visualizations reported different trust ratings depending on visual complexity.
% When designing our study, we deliberately chose a topic for the non-Covid visualization that would elicit fewer strong opinions. Hence, we intuitively find that when participants have no vested interest (e.g., for lack of political reasons), they are less likely to form strong opinions on trust in the visualization. 
% This phenomena is further supported by the qualitative responses: \textit{``Assuming I trust this source, I would not have reason to disbelieve it. However, because I have no knowledge of fungi and insect based crop diseases, I do not know how much this information matters...,"}. 

The second takeaway is that \textbf{ participants liked line charts more than bar charts and reported decreasing trust ratings as the line charts became more complex}. Measuring participants affective responses to the visualizations in addition to their overall trust rating allowed us to detect that when participants did not like a visualization (bar charts) varying the complexity had little to no effect on their trust ratings. Conversely, the line chart, which received higher affective ratings, captured a clear decrease in trust as the visualization became more complex. Qualitative responses that support this finding include \textit{``The different colored lines made it easy for me to figure out the data between the different age groups of vaccinated and unvaccinated people. It didn't see like there was anything deliberately misleading about it or anything lie that, so I trust it."} and \textit{``It looks like a graph that has reliable information on it, a line graph is used often."}

% We found that visual complexity significantly predicts the trust antecedent of \textit{cognitive clarity}.
% Since the \textit{cognitive clarity} trust antecedent also predicts overall trust in the visualization (Table \ref{table:trust_in_vis_data_stats}, column a), we can interpret the interaction effect of data topic and complexity on trust in the visualization as being facilitated by cognitive clarity. In other words, when the viewer has a vested interest in the visualization topic, they are more likely to trust a visualization that they find clear. 

The third takeaway is that \textbf{increased complexity in a visualization leads to decreased trust in the underlying data}. Given the results of our linear regression models for each of the data antecedents, we can assign this decreased trust as a result of decreased affective benevolence for increasingly complex visualizations. In other words, for simpler visualizations, participants find the data to be less biased and more trustworthy, leading to more overall trust in the data. Our qualitative responses suggest that simpler visualizations allowed the participants to validate their expectations of the data and infer that the data has a low bias: \textit{``Everything is displayed cleanly.  There is no evident bias"}.

% \vspace{-1mm}
\section{Discussion of Trust Framework}  

%Our framework provides significant value for future study design. 
Much existing trust research in data visualizations has measured trust using a single item that does not capture the underlying nuances and drivers of trust. By explicitly outlining two general dichotomies present in the formation of trust in visualizations, we provide guidance for measuring trust more comprehensively. In practice, this gives visualization researchers a theoretical framework on how to elicit trust in the visualization and underlying data that covers several antecedents of both cognitive and affective trust (e.g., clarity, accuracy, etc.). This also serves to lighten the burden of defining trust on the side of the participants.

By applying our framework to the case study on trust in visualizations, we were able to determine the specific trust antecedent which caused participants to trust simple visualizations more than complex ones. Our framework also gave us more granular insight into what aspects of our visual design most impacted trust. For example, changes in the chart type and complexity drove participants' assessment of visual clarity and their overall trust rating. These insights would not have been achievable with a single-item trust scale.
Additionally, by separating the antecedents into two categories for trust in visualization and trust in data, we found an interesting distinction. For visualization trust, the effect of complexity can be modulated by the chart type.  This phenomenon did not occur for data trust where complex visualization inhibited trust in the data regardless of the chart type. This distinction would not have been found via traditional methods of measuring trust in data visualizations. 

Aside from the trust antecedents outlined above, our framework also considers the individual differences that influence trust, as well as the behavioral outcomes that follow a trust judgment. By capturing and integrating individual differences into our regression models, we were able to account for them and uncover the driving role that visual design had on establishing trust. The self-reported behavioral outcomes were also captured and, in combination with the trust antecedents, revealed the importance of the \textit{accuracy}, \textit{aesthetics}, and \textit{clarity} of a visualization when using or sharing it with others.

We ultimately find that the development of trust is a multifaceted process wherein neither contextual nor individual factors independently determine the outcome. Rather, these factors jointly contribute to the development of trust. For example, trust in the organization responsible for the data or the visualization can significantly influence an individual's level of trust in the visualization, particularly when it comes to controversial and salient topics. Nevertheless, the impact of this trust can still be mitigated by the overall quality of the visualization. Moreover, the relationship between the different domains of trust (trust in the visualization, trust in the data, trust in science, interpersonal data) is shaped by the visual characteristics and topic of the visualization.

%Limitations:

% Our study items do not cover the entire framework
% We are missing items for affective visualization benevolence, 

% We should consider multiple items to cover some of the antecedents to trust 
%   for example: affective cues are captured only by ``I like this visualization ," and the "ugly - pretty" slider
%               we could add items like ``the visualization was aesthetically-pleasing," ``I enjoyed viewing this visualization," ``this visualization is artistic"

% We should also consider adding items with different valence to capture each antecedent from different angles
%   for example: cognitive clarity: ``I find it easy to understand this visualization" could be matched with ``This visualization was confusing"

% We should flesh out the initial impressions section with more effective items as well

% We only have two behavior items. 
% We could consider asking additional specific behavior items like: ``I am more likely to get vaccinated" 
%   or more general ones like: ``I am more likely to read similar visualizations in the future"

% \vspace{-2mm}
\section{Conclusions and Future Work}

Computer science trust research is still young, and most prior approaches to measuring trust have not been comprehensive. In this work, we provide a multidimensional conceptualization and operationalization of trust in visualization in a framework that builds on the trust literature and synthesizes existing work in the visualization field. The framework clearly outlines the antecedents to both cognitive and affective trust in visualizations and guides future research on measuring trust more comprehensively. 

We apply this framework to a large crowdsourced study that investigates the role of visual complexity on trust. Our findings provide initial evidence that elements such as visual complexity, viewers' vested interest in the topic, and their assessment of benevolence strongly influence their trust judgment.  

In future work, we intend to refine the case study scale items with the knowledge we gained. Some of these refinements include adding positive and negative items for each antecedent (e.g., we can measure \textit{cognitive clarity} via both ``I find it easy to understand this visualization" and  ``this visualization was confusing") and including more behavioral outcome items that target broader behavioral patterns (e.g., ``I am more likely to read similar visualizations in the future") as well as items specific to the dataset (e.g., ``I am more likely to get vaccinated").

We further intend to examine trust development in visualizations that intentionally elicit negative affective responses. Trust is more challenging to establish when dealing with trustees in a negatively valenced affective state \cite{dunn_feeling_2005}, as areas of the brain linked to negative emotions (e.g., fear of loss) are associated with a lack of trust or distrust \cite{dimoka_what_2010}. However, it is possible to cultivate trust in negative contexts. For instance, anxiety was found to decrease trust, whilst high-certainty emotions such as anger and guilt have no clear effect on trust \cite{myers_influence_2016}. Identifying the contextual factors that foster trust in negative visualizations presents a compelling direction for future research.

Other possible future research includes the implementation of trust games as a way of operationalizing the concepts in our trust framework.  Additionally, work done in belief updating \cite{kim_data_2018},  can inform trust measurements that compare a person's initial beliefs to their beliefs immediately after viewing the visualization. For capturing behavioral outcomes, a longitudinal study would support investigating the longer-term effects of trust. Ultimately, our work moves us towards a richer understanding of how visualizations can support or hinder trust formation, guiding the more intentional study and design of public-facing visualizations.

% Future work could try the following:
%   normalize based on interpersonal trust more explicitly (we only consider it in the linear regression model)
%   consider capturing trust as belief updating
%       beliefs after significant time has passed could also be interesting
% trust games?

%% if specified like this the section will be omitted in review mode
\acknowledgments{%
The authors wish to thank Barbara Kulaga for her role in developing the visualizations, and Bo Yun Park for her role in the original conception of the ``trust in science" experiment. This work was supported by the Harvard Data Science Initiative Trust in Science Fund, supported by Bayer, and partly funded by NSF awards IIS-1901030 and IIS-2237585.%
}

\bibliographystyle{abbrv-doi-hyperref}

\bibliography{VisTrust}

% \appendix % You can use the `hideappendix` class option to skip everything after \appendix

% \section{About Appendices}
% Refer to \cref{sec:appendices_inst} for instructions regarding appendices.

% \section{Troubleshooting}
% \label{appendix:troubleshooting}

% \subsection{ifpdf error}

% If you receive compilation errors along the lines of \texttt{Package ifpdf Error: Name clash, \textbackslash ifpdf is already defined} then please add a new line \verb|\let\ifpdf\relax| right after the \verb|\documentclass[journal]{vgtc}| call.
% Note that your error is due to packages you use that define \verb|\ifpdf| which is obsolete (the result is that \verb|\ifpdf| is defined twice); these packages should be changed to use \verb|ifpdf| package instead.

% \subsection{\texttt{pdfendlink} error}

% Occasionally (for some \LaTeX\ distributions) this hyper-linked bib\TeX\ style may lead to \textbf{compilation errors} (\texttt{pdfendlink ended up in different nesting level ...}) if a reference entry is broken across two pages (due to a bug in \verb|hyperref|).
% In this case, make sure you have the latest version of the \verb|hyperref| package (i.e.\ update your \LaTeX\ installation/packages) or, alternatively, revert back to \verb|\bibliographystyle{abbrv-doi}| (at the expense of removing hyperlinks from the bibliography) and try \verb|\bibliographystyle{abbrv-doi-hyperref}| again after some more editing.

\end{document}